\documentclass[aps.prb,twocolumn,showpacs,preprintnumbers,amsmath,amssymb]{revtex4}
\usepackage{mathrsfs}
\usepackage{graphicx}
\usepackage{amsmath}
\usepackage{color}

\begin{document}

\title{Rectification of Vortex Motion in a Circular Ratchet Channel}

\author{N.~S.~Lin$^{1}$, T.~W.~Heitmann$^{2}$, K.~Yu$^{2}$, B.~L.~T.~Plourde$^{2}$, and V.~R.~Misko$^{1}$}
\affiliation{$^{1}$Department of Physics, University of Antwerpen,
Groenenborgerlaan 171, B-2020 Antwerpen, Belgium \\
$^{2}$Department of Physics, Syracuse University, Syracuse, New York 13244-1130, USA}

\date{\today}

\begin{abstract}
We study the dynamics of vortices in an asymmetric (i.e., consisting
of triangular cells) ring channel driven by an external current $I$
in a Corbino setup. The asymmetric potential can rectify the motion
of vortices and cause a net flow without any unbiased external
drive, which is called ratchet effect. With an applied ac current,
the potential can rectify the motion of vortices in the channel and
induce a dc net flow. We show that the net flow of vortices in such
a system strongly depends on vortex density and frequency of the
driving force/current. Depending on the density, we distinguish a
``single-vortex" rectification regime (for low density, when each
vortex is rectified individually) determined by the potential-energy
landscape inside each cell of the channel (i.e., ``hard" and ``easy"
directions of motion) and ``multi-vortex", or ``collective",
rectification (high density case) when the interaction between
vortices becomes important. The frequency of the driving ac current
determines a possible distance that a vortex could move during one
period. For high frequency current, vortices only oscillate in the
triangular cell. For low frequency, the vortex angular velocity
$\omega$ increases nearly linearly until the driving force reaches
the maximum friction force in the hard direction. Furthermore, the
commensurability between the number of vortices and the number of
cells results in a stepwise $\omega-I$ curve, which means that the
average angular velocity $\omega$ of the vortices is a discontinuous
function of the applied current. Besides the ``integer" steps, i.e.,
the large steps found in the single vortex case, we also found
fractional steps corresponding to fractional ratio between the
numbers of vortices and triangular cells. The principal and
fractional frequencies for different currents are found, when the
net flow of vortices reaches the maximum that is proportional to the
frequency when the density of vortices is low.
We have performed preliminary measurements on a device
containing a single weak-pinning circular ratchet channel in a
Corbino geometry and observed a substantial asymmetric vortex
response.
\end{abstract}
\maketitle

\section{Introduction}
A net flow of particles under unbiased external fluctuations/drive
due to an asymmetric potential, which is called ratchet effect, has
received much attention during the last decades. The transport and
dynamical properties of particles on asymmetric potential have been
widely studied, e.g., in physics and biology
\cite{SMadd1,SMadd2,2PRB2005Peeters,mod,7PRB2005Nori,10PRB2007Kes,13PRL1999Nori,IEEE2009Plourde}.
Random motion of particles can be rectified in such an asymmetric
system, which can be used for, e.g., controlling particles motion,
separating different types of particles (i.e., molecular sieves),
for both underdamped and overdamped particles \cite{7PRB2005Nori}
and for molecular motors \cite{mod}. Vortices in a type II
superconductor often (e.g., for magnetic field close to $H_{c1}$)
can be treated as classical overdamped ``particles". Most of the
experiments on vortex motion rectification used arrays of asymmetric
pinning sites (e.g., nanoengineered antidots or triangular magnetic
dots/inpurties) to create an asymmetric potential, which rectifies
the motion of vortices
\cite{IEEE2009Plourde,3Nature2006,4Scinece2003,8PRB2005Vincent,9PRL2005Moshchalkov,12PRB2010Mashchalkov,11PRB2010Plourde,highTC2004}.
The rectified vortex motion was directly observed in experiments by
imaging vortices via Lorentz microscopy \cite{PRL2005Tonomura}.
Periodic arrangement of point defects of a gradual density or
periodic square array of ferromagnetic dots of decreasing size,
i.e., varying the density of pinning sites or the size of pinning
sites, were shown to result in a ratchet potential
\cite{6PRL2001Nori,16PRL2007Reichhardt}. When vortices are trapped
by pinning sites, the repulsive vortex-vortex interaction creates a
higher energy barrier near the area with higher density of pinning
sites. Therefore, an asymmetric potential can be created by the
gradient of the density of pinning sites \cite{6PRL2001Nori}. Even
without spatial asymmetry (i.e., without any asymmetric
walls/boundaries or asymmetric pinning sites), the motion of
vortices still can be controlled by time-asymmetric driving force
\cite{15ref1Nori,15ref2Marchesoni,15NMat2006Nori}. Due to the
possibility of controlling their motion, the dynamical behavior of
vortices in such systems has attracted considerable interest. A
series of elastic and plastic vortex flow phases were found
\cite{Nori_Reichhardt,PRL2006PRB2007misko,1PRB2010Reichhardt}.
Besides the liquid-like and solid lattice phase, vortex motion also
revealed a jamming behavior
\cite{1PRB2010Reichhardt,14PhysC2010Reichhardt}. When the density of
vortices is changed, the vortex flowing direction can change to the
opposite
\cite{3Nature2006,4Scinece2003,16PRL2007Reichhardt,17PRB2007Reichhardt},
which means vortices can drift in either the ``hard" direction or
the ``easy" direction of the ratchet, depending on the vortex
density. By controlling the motion of vortices, it is possible to
remove vortices or reduce the vortex density by using a combination
of two opposite oriented ratchet arrays \cite{Nature1999}. The order
of vortices and commensurability between vortices and cells also
play an important role in vortex dynamics
\cite{16PRL2007Reichhardt,1PRB2010Reichhardt,17PRB2007Reichhardt,11PRB2010Plourde}.
In two-dimensional (2D) ratchets, the dynamics in the transverse
ratchet was also studied in theory \cite{7PRB2005Nori} and in
experiments \cite{TransRatchetAPL2007,TransRatchetAPL2008}.

In the present paper, we study the dynamics of vortices in a
circular channel formed by asymmetric triangular (funnel) cells
(TCs) (see Fig. \ref{fig:geo}) [note that earlier this approach 
to form asymmetric channels in experiment (i.e., using weak-pinning 
channels) was employed in a stripe geometry~\cite{10PRB2007Kes}].
Due to the radially flowing current
in a Corbino setup, the driving force is not uniform inside the cell
which is different from linear ratchet channels. Such a geometry
(i.e., asymmetric in the azimuthal direction and in the radial
direction), as will be shown, leads to a specific dynamical
behavior, for example, a vortex located near the inner corner of a
TC (which is closer to the center), experiences a stronger driving
force and moves to the next TC while a vortex located near the outer
corner of TC does not move. The circular geometry of the ratchet
channel is convenient for studying commensurability and step-motor
(phase locking) behavior
\cite{12PRB2010Mashchalkov,1PRB2010Reichhardt}. We analyze in detail
low and high density regimes of rectification, i.e.,
``single-vortex" and ``multi-vortex" regimes. We demonstrate that
the mechanism of rectification are qualitatively different for these
two cases.
In addition, we have performed preliminary measurements using 
a single nanofabricated weak-pinning ratchet channel of a-NbGe 
with strong-pinning NbN channel edges in a Corbino set-up, and 
we observed a substantial asymmetric vortex response.

This paper is organized as follows. First in Sec.~II, the model of
our systems will be presented. Then in Sec.~III, we will show how
the density of vortices and the frequency of driving current
influences the dynamical behavior of vortices. Commensurability
effects of vortex density and of the frequency of current will be
discussed in Sec.~IV.
We present the measurements of a vortex ratchet in a Corbino geometry in Sec.~V.
Finally, the conclusions will be given in Sec.~VI.

\section{Model and simulation}



\begin{figure}[tb]
\centering
\includegraphics[width=0.9\linewidth]{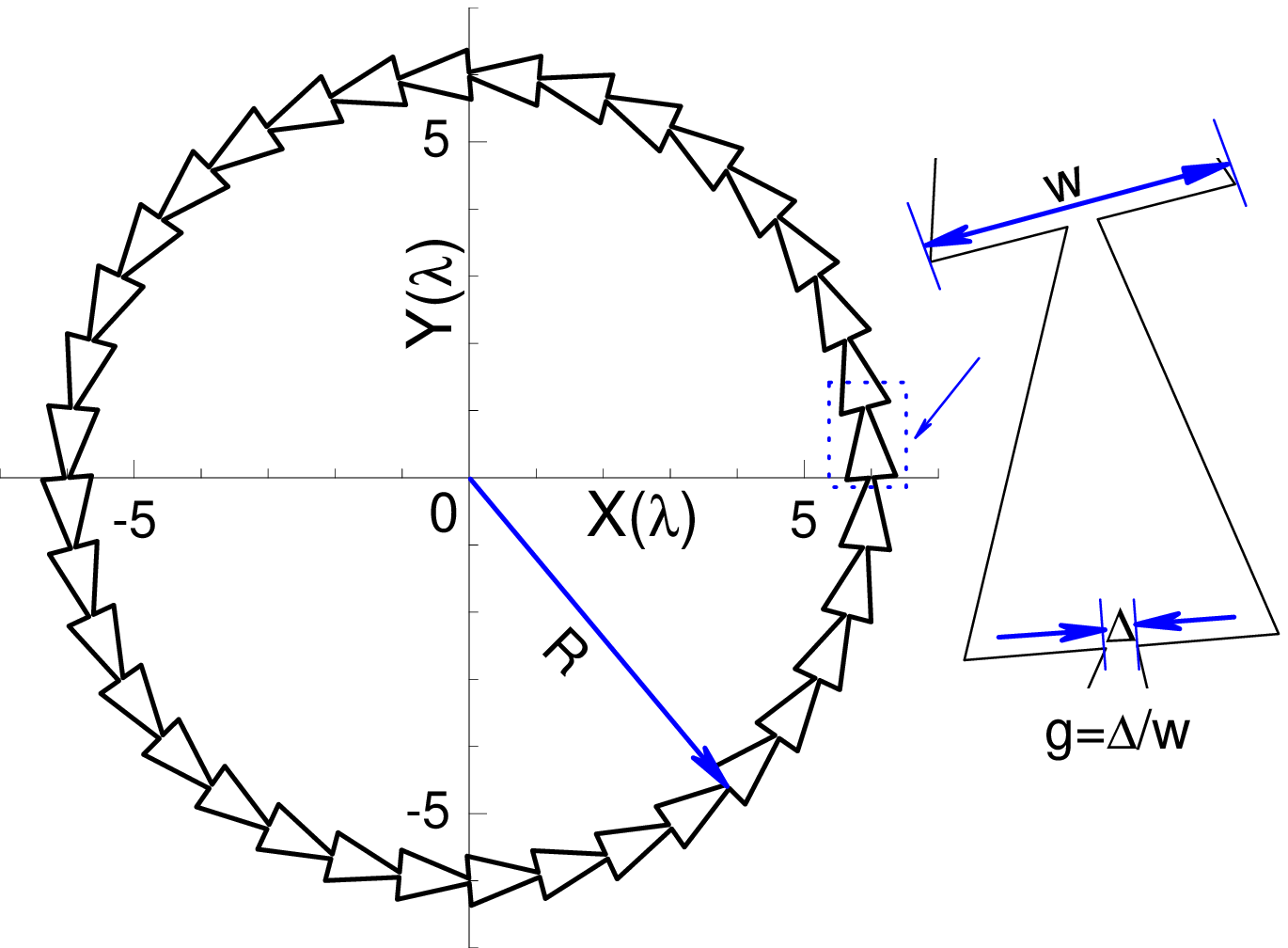}\\
\caption{
(Color online)
The geometry of the system. The widths of the wide part $\omega$ and
narrow part $\Delta$ are shown in the figure, where $g=\Delta/w$ is
the ratio between the two parts. Here $g=0.1$, the radius of the
channel $R=6\lambda$ and the channel is constructed by 36 TCs, i.e.,
$N=36$.}
  \label{fig:geo}
  \includegraphics[width=0.9\linewidth]{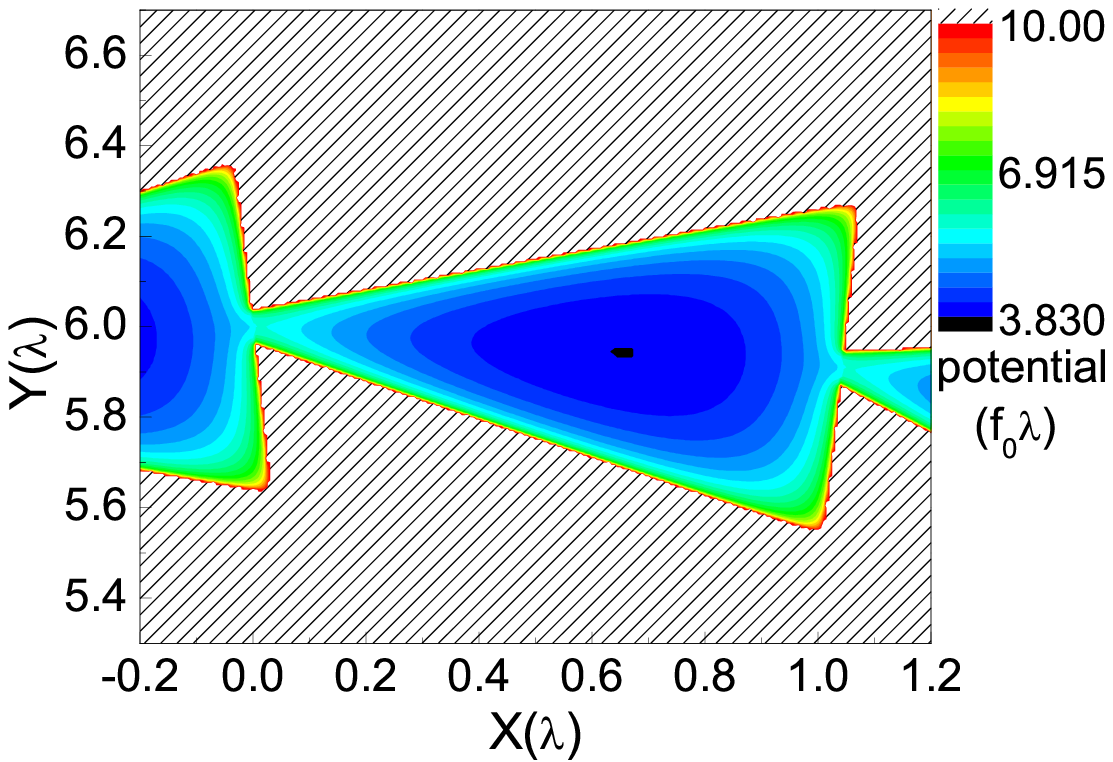}\\
  \caption{
(Color online)
The contour map of the modeled potential. The lowest
  potential in a TC is close to the geometry center of the triangle
  and the potential near the boundary is high enough to prevent
  vortices from escaping the TC.}
  \label{fig:potential}
\end{figure}

We consider a ring-like weak-pinning channel constructed of $N$
partially overlapping TCs as shown in Fig. \ref{fig:geo}. In our
simulations, the radius of the ring $R$ is typically set as
$6\lambda$, where $\lambda$ is the magnetic field penetration depth,
the wider part of the channel (i.e., the base of TCs)
$w=0.75\lambda$, and the width of the narrow part $\Delta$ (the
neck) is typically 0.1 of the wider part (i.e., the ratio
$g=\Delta/w=0.1$). We also performed simulations in a channel with a
wider neck part, e.g., $g=0.15$, for comparison. The weak-pinning
channel (where vortices can move freely) is surrounded by a
strong-pinning superconducting material \cite{Plourde2008} which is
modeled by a medium where vortices cannot move. An external current
$I$ radially flows from the center of the disk to the edge,
resulting in the density of current $J(\rho)\sim I/\rho$. Therefore,
the closer a vortex to the center of the disk, the stronger the
Lorentz force that acts on it. The driving force due to the radial
current is
\cite{Marchetti2002,Crabtree1999,DynamicHTSC,lgCorbinoMiguel,smCorbinoMisko,PRLLinMisko}
\begin{equation}\label{eq:drvf}
\boldsymbol{f}_i^d=\frac{\Phi_0 I}{2\pi\rho_i d}\ \hat {\theta} =
\frac{f_0 I_0}{\rho_i}\ \hat {\theta},
\end{equation}
where $\hat{\theta}$ is the unit vector in the azimuthal direction
in the disk plane, $f_0=\Phi_0^2/(2\pi\mu_0 \lambda^3)$ is the unit
of force, $I_0=\mu_0\lambda^2 I/(\Phi_0 d)$ is the dimensionless
driving current, and $\Phi_0$ is the flux quantum. We perform
Langevin-type molecular dynamics (MD) simulations and numerically
integrate the overdamped equations of motion
\cite{smCorbinoMisko,PRLLinMisko,fvvBae,Grigorieva2007}:
\begin{equation}\label{eq:motion}
\eta \frac{d\boldsymbol{r}_i}{dt}=\boldsymbol{f}_i
\end{equation}
with
\begin{equation}
\boldsymbol{f}_i=\sum_{j}\boldsymbol{f}_{ij}^{vv}+\boldsymbol{f}_i^{\,T}
+\boldsymbol{f}_i^d+\boldsymbol{f}_i^b,
\end{equation}
where $\eta$ is the dimensionless viscosity coefficient which is set here to unity. 
We note that in practice the viscosity coefficient 
\cite{Bardeen65} varies between different superconductors. 
As an example, parameters for a-NbGe, a typical weak-pinning 
material, correspond to 
$\eta \approx 10^{-8}$~Ns/m$^{2}$ \cite{babic}. 
Using this value of $\eta$ in our calculations 
results in typical maximum values of vortex linear velocity 
$v \approx 10^{2}$~m/s (for a 1$\mu$m-thick film), 
which approaches, but does not exceed, the 
Larkin-Ovchinnikov critical velocity for this material 
\cite{LO,klein,babic}.

The vortex interaction $\boldsymbol{f}_{ij}^{vv}$ is
described by a first order of the modified Bessel function of the
second kind $K_1(r_{ij}/\lambda)$, and the thermal force obeys
\begin{equation}
\langle \boldsymbol{f}_i^T(t)\rangle=0,
\end{equation}
and
\begin{equation}
\langle\boldsymbol{f}_i^T(t)\boldsymbol{f}_j^T(t')\rangle=2\eta
k_BT\delta_{ij}\delta(t-t').
\end{equation}

To model the vortex-boundary interaction, we assume an infinite
potential wall at the boundary (i.e., vortices cannot leave the channel) which decays inside the triangular cell
with the same dependence on position
as the vortex-vortex interaction potential.
The total interaction of a vortex with the channel boundaries
is calculated by integrating the vortex-wall interaction potential
over the geometrical boundary of the channel.
The resulting potential due to the boundary is shown in
Fig.~\ref{fig:potential}, and the vortex-boundary force
$\boldsymbol{f}_i^b$ is directly calculated from the potential.

In our simulations, we first set $T>0$, when no current is applied, and then gradually decrease temperature to let the system to relax to the ground state.
Then we set $T=0$ and apply an external driving, i.e., an ac current resulting in oscillating Lorentz force with frequency $\nu$ and amplitude $I_{0}$ that acts on vortices, to study the dynamics of the system.

\section{Rectification of vortex motion}

\subsection{Density of vortices}

The vortex density, i.e., the number of vortices per TC, plays an
important role in the dynamics of the system. For very low vortex
density, the vortex-vortex interaction is small as compared to the
interaction with the boundary, and therefore it can be neglected.
With increasing the vortex density, the interaction between vortices
becomes important. Therefore, we roughly distinguish two regimes,
i.e., a single-vortex regime and a multi-vortex regime,
respectively, corresponding to low and high vortex density.

\begin{figure*}[tb]
\begin{minipage}[c]{0.48\linewidth}
\centering
\includegraphics[width=0.85\textwidth]{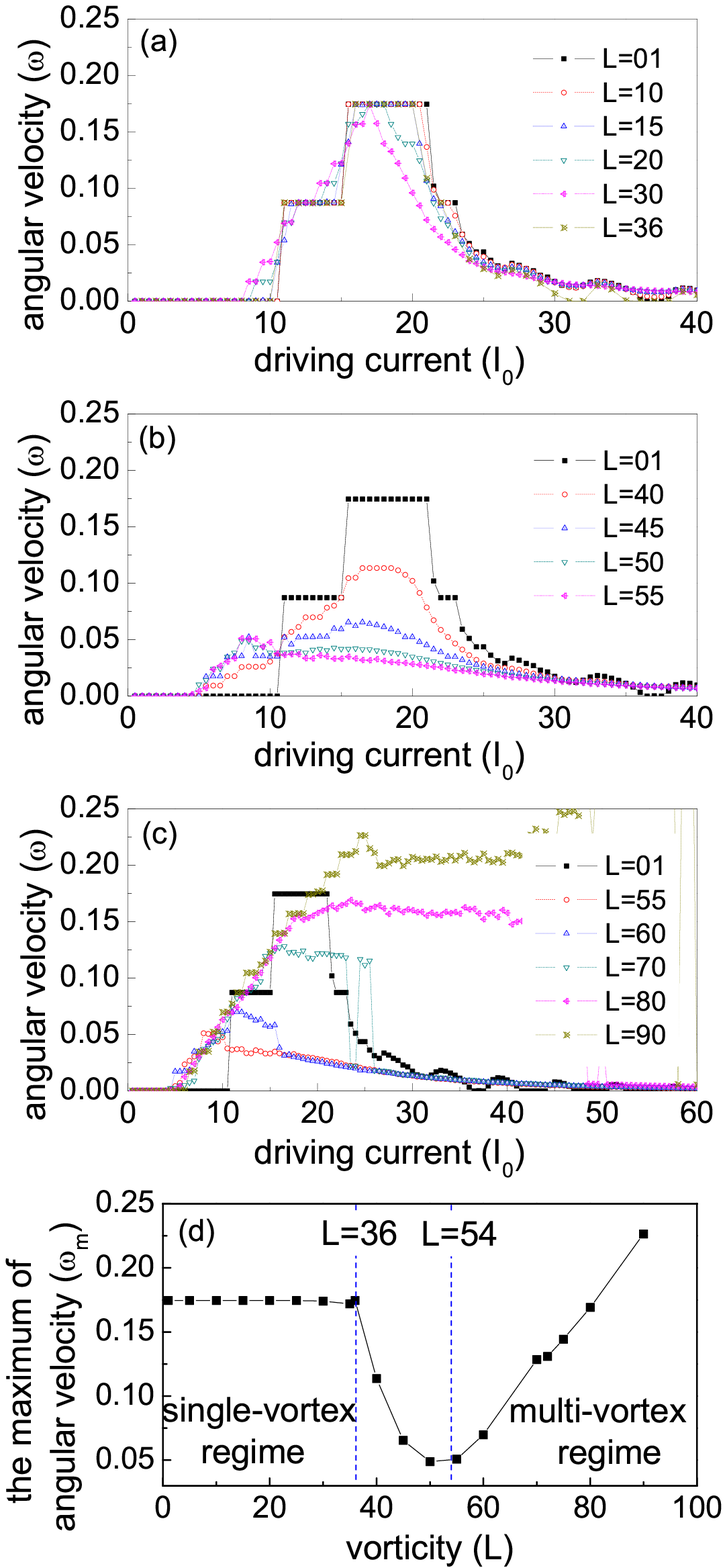}
\caption{
(Color online)
The $\omega-I_0$ curves for different density of vortices:
(a) the density of vortices is low, $L=1-36$ (in the single-vortex
regime), the angular velocity reaches the same maximum; (b) the
density increases, $L=40-55$, the maximum of the angular velocity
decreases; (c) for further increasing density, i.e., high density of
vortices, $L=55-90$, the maximum starts to increase and the critical
value of the driving current, when the angular velocity starts
decreasing, becomes larger; (d) the maximum of the angular velocity
$\omega_m$ first remains the same until the number of vortices
$L>36$. Then the maximum $\omega_m$ first decreases, but for $L>55$
it starts to increase. }\label{fig:iw_incom}
\end{minipage}
\hfill
\begin{minipage}[c]{0.48\linewidth}
\includegraphics[width=0.54\textwidth]{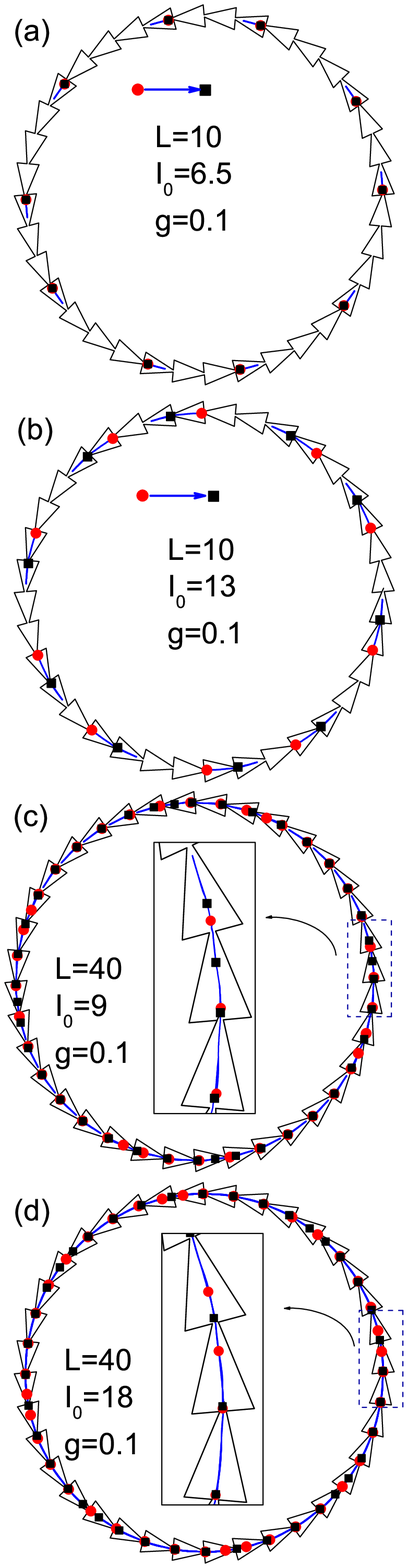}
\caption{
(Color online)
Trajectories of vortex motion, in the case of the low
density of vortices ($L=10$)  (a), (b),  and in the intermediate
case ($L=40$) (c), (d), for different values of driving current:
$I_0=$ 6.5 (a), 13 (b), 9 (c), and 18 (d). The initial positions of
the vortices are marked by gray circles and the final positions
after one period of ac current are marked by black squares. The
trajectories of vortices, which are plotted by solid lines, are
nearly circular no matter that the current is large or small.} \label{fig:trace_low}
\end{minipage}
\end{figure*}

In order to characterize the dynamical behavior of the system of vortices, 
we calculate the net angular velocity of each vortex and take average 
over all the vortices for, e.g., 100 ac periods.
The resulting average angular velocity (called further ``angular velocity'') 
is denoted as $\omega$ which is analyzed for different parameters of the 
system and drivings.
%
We note that angular velocity can be related in a straightforward way to the 
flux-flow voltage \cite{Plourde2008,DynamicHTSC} that one would typically measure in an experiment. 

One system contains $L$ vortices and $N$ triangles in a circular
chain, e.g., in our simulations we take $N=36$.
If there is less than one vortex per cell, i.e., $L<36$ in our
system (see Fig. \ref{fig:geo}), then the density of vortices is low
enough and one can neglect the interaction between vortices. One can
imagine that all the vortices are far away from each other and
weakly interact with other vortices but they are strongly influenced
by the ratchet potential induced by the boundary. Therefore, the
dynamical behavior in the low density case is similar to that of the
system with just one vortex [shown in Fig. \ref{fig:iw_incom} (a)],
which is considered in the single-vortex regime. Thus, for $L=1-36$,
the maximum of the angular velocity is the same value [see Fig.
\ref{fig:iw_incom} (a)]. When the density increases and therefore
the interaction between the vortices becomes appreciable, the
maximum of the angular velocity $\omega_m$ first decreases to 0.05
for $L=55$ [see Fig. \ref{fig:iw_incom} (b)] and then starts to
increase [see Fig. \ref{fig:iw_incom} (c)] when the system turns to
the regime of high vortex density.

\begin{figure*}[tb]
\begin{minipage}[t]{0.32\linewidth}
\centering
\includegraphics[width=\textwidth]{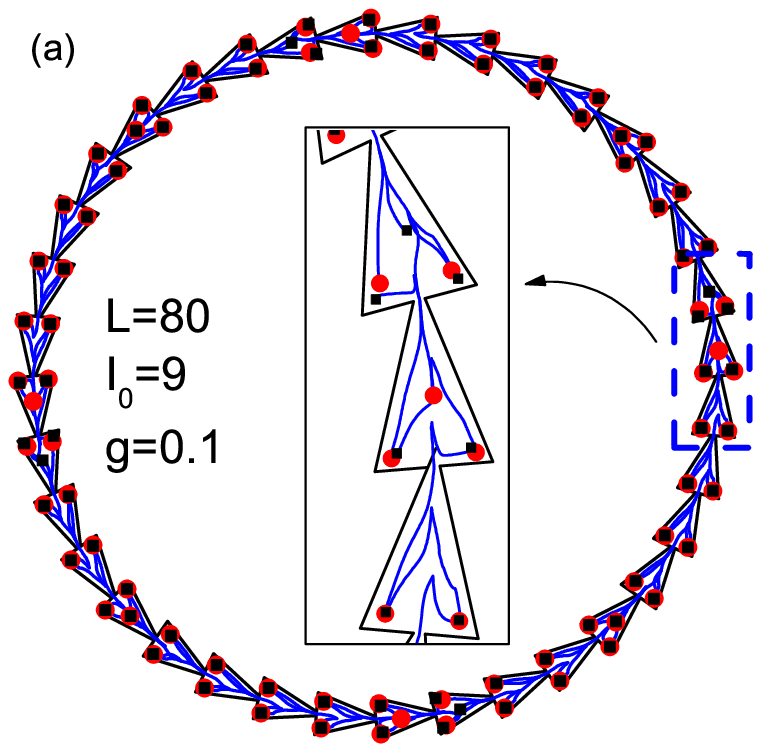}
\end{minipage}
\hfill
\begin{minipage}[t]{0.32\linewidth}
\centering
\includegraphics[width=\textwidth]{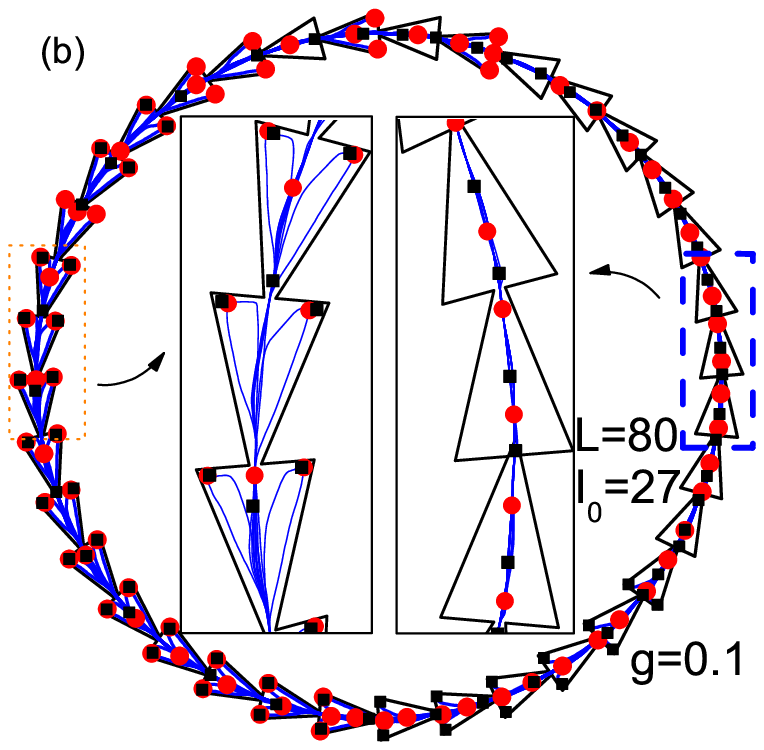}
\end{minipage}
\hfill
\begin{minipage}[t]{0.32\linewidth}
\centering
\includegraphics[width=\textwidth]{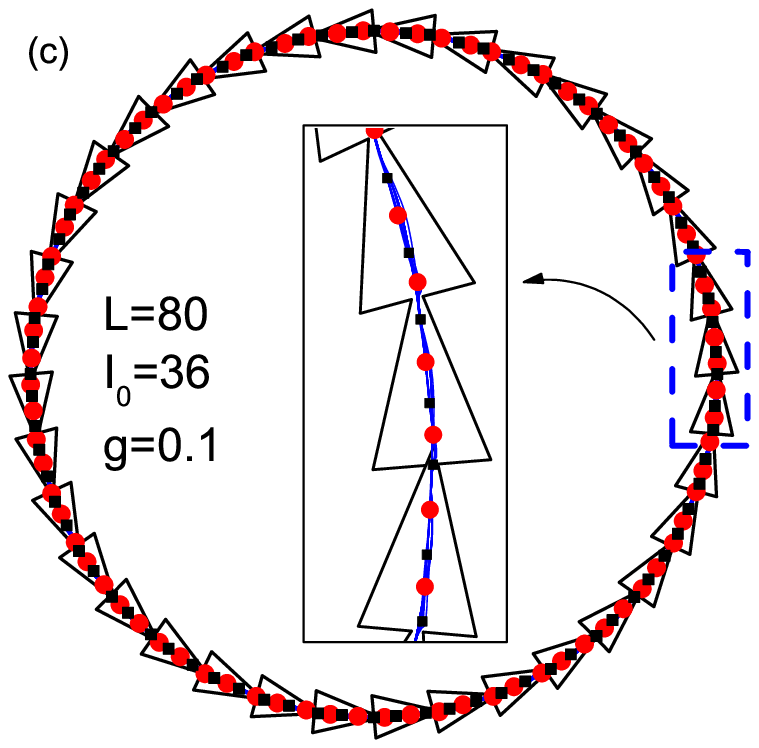}
\end{minipage}
\caption{
(Color online)
Trajectories of vortex motion during one ac period in the
case of high density of vortices ($L=80$), i.e., the multi-vortex
regime, for different values of the driving current, $I_0=$ 9 (a),
27 (b), and 36 (c). When the current is in a certain range
$I_0=5-19$ [see Fig. \ref{fig:iw_incom} (c)], the vortices move
along 2D trajectory [e.g., $I_0=9$, shown in (a)] which is different
from the circular 1D trajectories in the low density case.
With increasing current, more vortices start to move along circular trajectories and thus 2D trajectories finally turns to a 1D circle [e.g., see (c)].
During the transition from the 2D to 1D motion, the
distribution of vortices becomes inhomogeneous [shown in
(b)].}
\label{fig:trace_high}
\end{figure*}

To understand this non-monotonic behavior of the maximum of the
angular velocity $\omega_m$ [shown in Fig. \ref{fig:iw_incom}(d)],
we studied the trajectories of moving vortices. When the system is
in the single-vortex regime, where $L\leq36$, all the vortices move
along a nearly circular trajectory no matter whether the applied
current is small or large, which can be considered as a
one-dimensional (1D) motion (shown in Fig. \ref{fig:trace_low}). For
low driving currents, each vortex oscillates near its initial
position inside a TC [e.g., see Fig. \ref{fig:trace_low} (a)], and
when the driving force reaches some critical value, all the vortices
move with a net angular velocity $\omega$ in the easy direction
[e.g., see Fig. \ref{fig:trace_low} (b)]. However, in the
multi-vortex regime, i.e., the high density case (e.g., $L=80$), the
motion of vortices is not 1D any more. The ac current drives
vortices to pass through the narrow part (i.e., neck) of the channel
in the easy direction and when current is alternated, some of
vortices are forced to move into the corner by the others and
``freeze" thus blocking the motion in the hard direction [shown in
Fig. \ref{fig:trace_high}]. Due to the asymmetry in the radial
direction and radially decreasing current density in the Corbino
setup, a vortex near the inner corner, which is closer to the center
of the disk, moves faster than the one near the outer corner.
Therefore, at a specific value of the current [e.g., at $I_0=9$, as
shown in Fig. \ref{fig:trace_high}(a)], the vortex near the inner
corner has a larger angular velocity. When the current increases,
more and more vortices move along the circular trajectory and the
distribution of vortices becomes strongly inhomogeneous [shown in
Fig. \ref{fig:trace_high}(b)].
With further increasing current, this 2D trajectory becomes narrow and all the vortices move along the circle, i.e., the 2D motion becomes 1D [shown in Fig.~\ref{fig:trace_high}(c)].
With increasing density, more and more vortices remain in the area near the corners of TCs and block other vortices moving in the hard direction.
Therefore, the maximum of vortex angular velocity increases. Even in the descending-velocity region of the single-vortex regime [see e.g., when $I_0>21$ in Fig.~\ref{fig:iw_incom}(c)], the angular velocity is still large.

The maximum of the angular velocity $\omega_m$ for different
vorticities $L$ is shown in Fig. \ref{fig:iw_incom}(d). In the
single-vortex regime, where there is less than one vortex per TC,
$\omega_m$ remains almost the same value because of the weak
interaction between vortices. When $L$ becomes larger, there are
more than one vortex per TC and the repulsive vortex-vortex
interaction in the same TC makes the motion in either easy or hard
directions much easier. Therefore, $\omega_m$ first decreases when
$L>36$. However, for $L\geq55$, the maximum $\omega_m$ starts to
increase. The reason for this behavior is the transition from 1D
motion to 2D motion. When vortices move along 2D trajectories (see
Fig. \ref{fig:iw_incom}), the repulsive interaction between vortices
still ``helps" a vortex to move to the next TC in the easy direction
but blocks the vortex motion in the hard direction until applying a
large enough driving current.

The radial asymmetry facilitates the motion of vortices from
inner/outer corners (driven by stronger/weaker force) one by one.
This is different from the case of a linear ratchet channel where
two vortices in the corners are driven by the same force and, as a
result, arrive simultaneously at the neck region leading to jamming.
In an asymmetric channel, the symmetry is broken, and jamming occurs
only for rather large driving force.

\subsection{Frequency dependence}

\begin{figure*}[tb]
\centering
\includegraphics[width=0.90\textwidth]{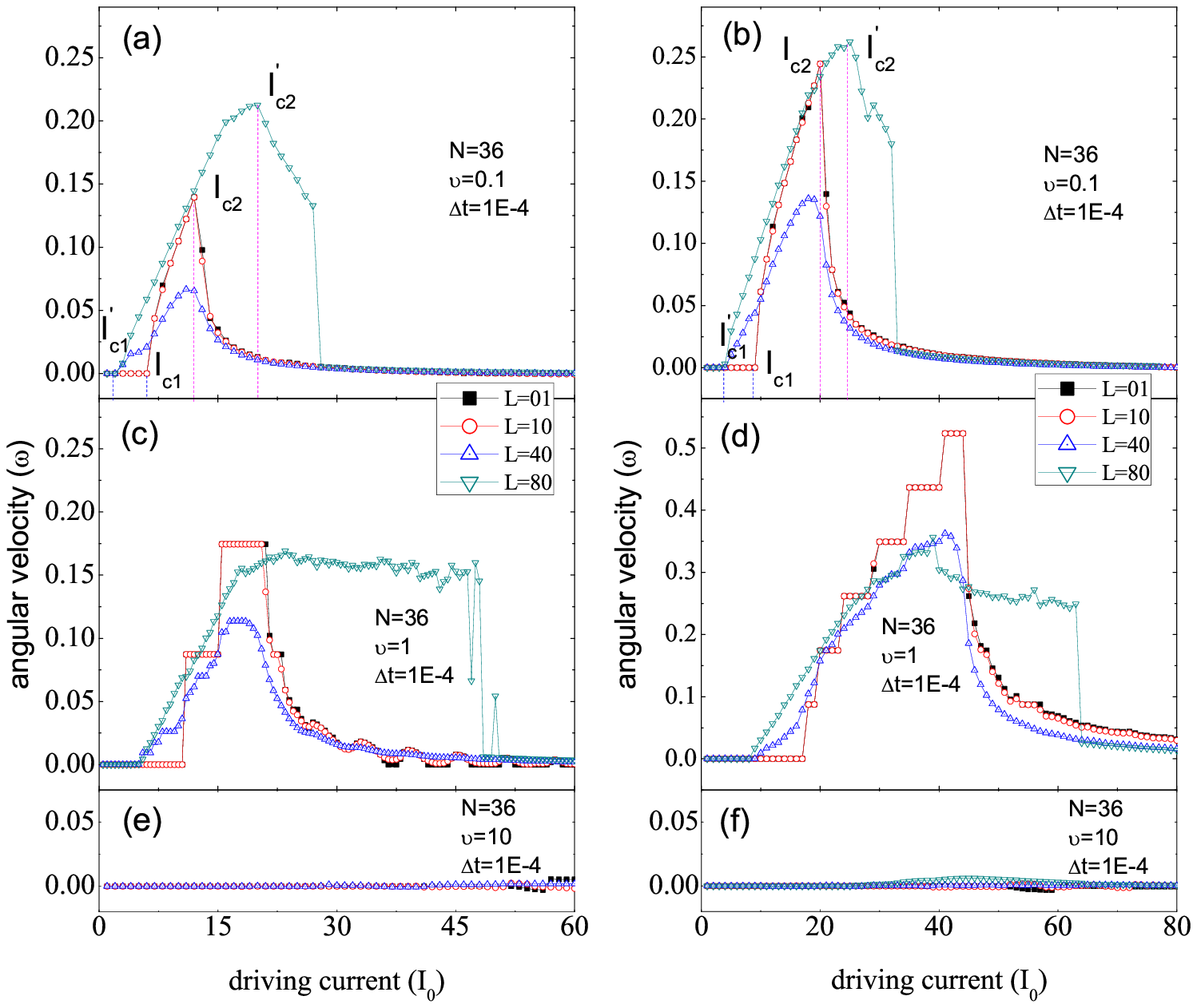} 
\vspace{-1cm}
\caption{
(Color online)
The $\omega-I_0$ curves for different frequencies:
$\nu=0.1$ (a), (b), $\nu=1$ (c), (d), and $\nu=10$ (e), (f). (a),
(c), and (e) are in the system with a narrower neck, $g=0.1$, and
(b), (d), and (f) are in the system with a wider neck $g=0.15$. The
scale of $\omega$ in panel (d) is different from the others. For low
frequency current (e.g., $\nu=0.1$) the ratchet effect is clearly
seen [as shown in (a) and (b)]. In this case, the angular velocity
first increases continuously to the maximum and then decreases to
zero. For intermediate frequency $\nu=1$, besides the ratchet
effect, the effect of commensurability, i.e., discontinuity in the
angular velocity $\omega(I_0)$ (see Sec. IV), has a clear influence
on the $\omega-I_0$ curve [e.g., see (c) and (d) for $L=1$ or
$L=10$]. For high frequency current (e.g., $\nu=10$), vortices are
confined in their initial TCs and only oscillate in a single
potential well. Therefore, the effect of the periodic ratchet
potential disappears [shown in (e) and (f)].}\label{fig:iw_gap}
\end{figure*}

The dynamical behavior of vortices in the considered ratchet system
is also strongly influenced by frequency of the ac current. In this
subsection we study frequency dependence of the rectified vortex
motion. For different frequencies $\nu$, the $\omega-I_0$ curves are
shown in Fig. \ref{fig:iw_gap}.

When the frequency is low, the dynamical behavior is similar for
different vortex densities.
As the drive amplitude $I_0$ is increased, the angular velocity first increases, reaches a maximum, and then decreases to zero. The curves, especially for $L=1$, in
Fig. \ref{fig:iw_gap}(a) are similar to the analytic result in a
ratchet potential (e.g., see Fig. 2 in Ref. \cite{Nature1999}). In
the case of $L=1$, the first critical value of current $I_{c1}$
[shown in Fig. \ref{fig:iw_gap}(a)] corresponds to the maximum
friction in the easy direction $f^+_{m}$ and the second one $I_{c2}$
corresponds to the case when the driving force reaches the maximum
friction in the hard direction $f^-_{m}$. When $L=40$, the first
critical value of current decreases to $I_{c1}^\prime$, which means
vortices are easier to move in the easy direction. If the density
increases further, $I_{c1}^\prime$ decreases and $I_{c2}^\prime$
increases, for $L=80$ [shown in Fig. \ref{fig:iw_gap}(a)].
Therefore, the interaction between vortices allows vortices to move
even easier in the easy direction and harder in the hard direction.

For an intermediate frequency [e.g., $\nu=1$ as shown in Fig.
\ref{fig:iw_gap}(c)], when the distance a vortex moves during one
period is comparable to the size of a TC, the dynamical behavior of
vortices will depend not only on the driving force but also on the
vortex density. This is explained by the fact that the confinement
force due to the boundary and the vortex-vortex interaction are also
comparable. In this case we obtain different dynamical behavior in
single-vortex and multi-vortex regimes which were discussed in the
previous subsection. Due to the commensurability between the numbers
of TCs and vortices, a step structure of the $\omega-I_0$ curve is
revealed in the low density case (e.g., $L=1$ and $L=10$), which
will be discussed in more detail in Sec. IV.

As one can predict, the angular velocity becomes zero [see Fig.
\ref{fig:iw_gap}(e)] when the frequency is high enough because a
vortex oscillates near its initial position inside a cell and thus
its motion cannot be rectified. Therefore, the variation of the
angular velocity is much smaller than that in the low/intermediate
frequency case. Each vortex is localized in a specific TC, i.e., the
vortex is only influenced by a single potential well but not by the
ratchet potential.

For comparison, we also calculate $\omega-I_0$ curves for the case
when a TC has a wider connection part ($g=0.15$) [shown in Fig.
\ref{fig:iw_gap}(b), (d), (f)]. The function $\omega(I_0)$ in
general shows a similar behavior as for $g=0.1$. However, for the
intermediate frequency ($\nu=1$) of the applied current, we obtain
more steps in $\omega-I_0$ curve in the case of $L=1$ [shown in Fig.
\ref{fig:iw_gap}(d)] than that for $g=0.1$ [shown in Fig.
\ref{fig:iw_gap}(c)]. This relates to the commensurability effect
that will also be discussed in Sec. IV.

\section{Commensurability effect}

\subsection{Commensurability of vortex density}

\begin{figure}[tb]
\centering
\includegraphics[width=0.48\textwidth]{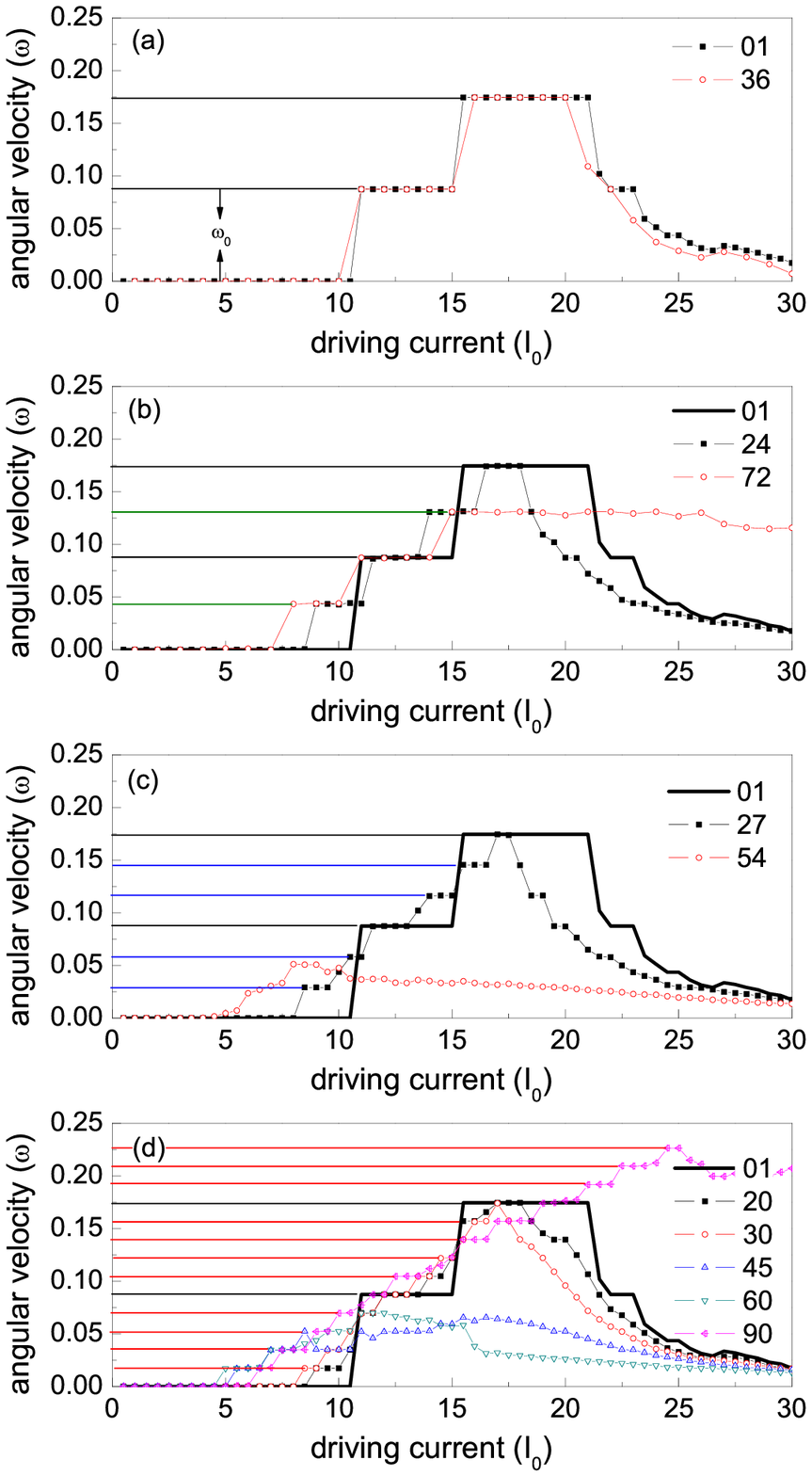}
\caption{
(Color online)
The $\omega -I_0$ curve for
a channel with $g = 0.1$ and different $k$: $k=$ 1 (a), 2 (b),
3 (c), and 5 (d). $L$ is the vorticity, i.e., the number of
vortices. $N$ is the number of cells in the circular chain. If
$L/N=1/m$, e.g., $L=1, 36$, the $\omega-I_0$ curve shows several
steps and the height of each integer step is $\omega_0$. When
$L/N=k/m$ (k=2, 3, 5), the height of each fractional step is
$\omega_0/k$ [e.g., as shown in (b) (c) and (d)].}\label{fig:iw_com}
\end{figure}

\begin{figure*}[tb]
\begin{minipage}[c]{0.32\linewidth}
\centering \caption{
(Color online)
(a) The configuration of 24 vortices after 1000
periods of oscillations, $t=t_i$. The moving trajectories of
vortices in one period of ac current, i.e., from $t=t_i$ to
$t=t_i+T$, is shown for $I_0=10$ (b), 13 (c), and 15 (d). The insets
of (a)-(d) show the vortex distribution in a unit cell. For $L=24$,
a type-A vortex is the vortex in a triangular cell (TC) [e.g., TC C2
in the inset of (a)] that has a neighbor empty TC in the easy
direction [e.g., TC C3 in the inset of (a)], i.e., a TC without
vortex, while a type-B vortex is in a TC [e.g., TC C1 in the inset
of (a)] that neighbors a TC with type-A vortex inside. When the
current drives vortices in the easy direction (e.g., $I_0=10$), two
types of vortices move in different ways. Type-A vortices move from
C2 to C3 but type-B vortices do not move to C2 due to a larger
repulsive interaction in the easy direction (b). When $I_0=13$ (c),
$\omega=\omega_0$ is the same as the integer step in $\omega-I_0$
curve [see Fig. \ref{fig:iw_com}(b)]. The type-A (type-B) vortex
moves over one TC and remains type-A (type-B) in the end of the
period. If the current increases further, e.g., $I_0=15$, type-A and
type-B vortices move with different net angular velocities and after
every period they switch their type [shown in
(d)].}\label{fig:typeab}
\end{minipage}
\hfill
\begin{minipage}[c]{0.32\linewidth}
\centering
\includegraphics[width=\textwidth]{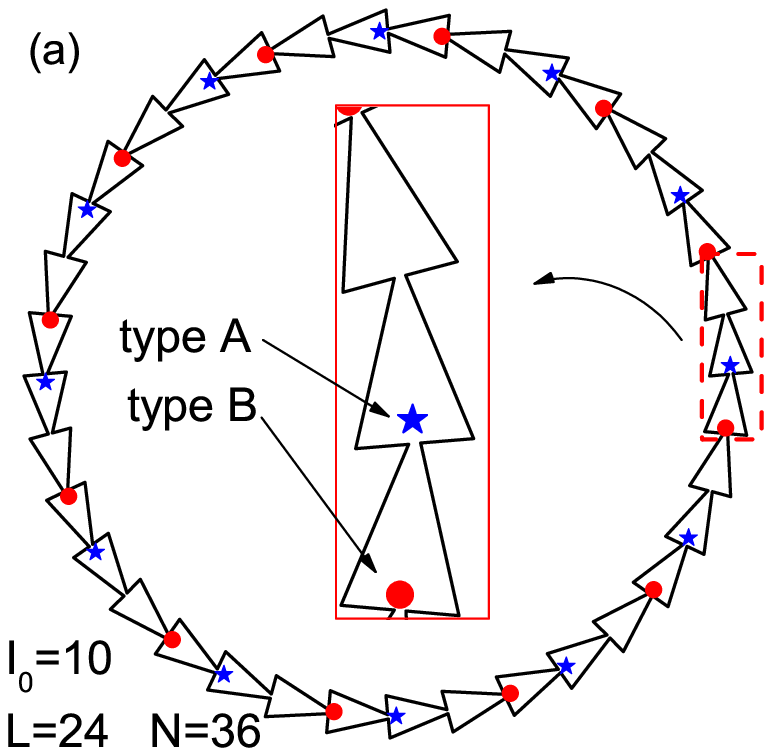}
\includegraphics[width=\textwidth]{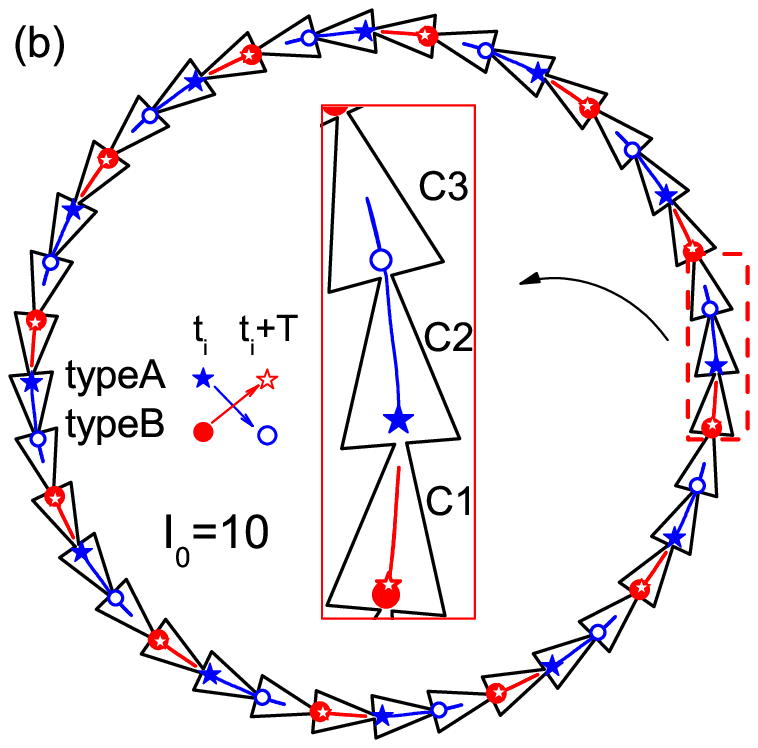}
\end{minipage}
\begin{minipage}[c]{0.32\linewidth}
\centering
\includegraphics[width=\textwidth]{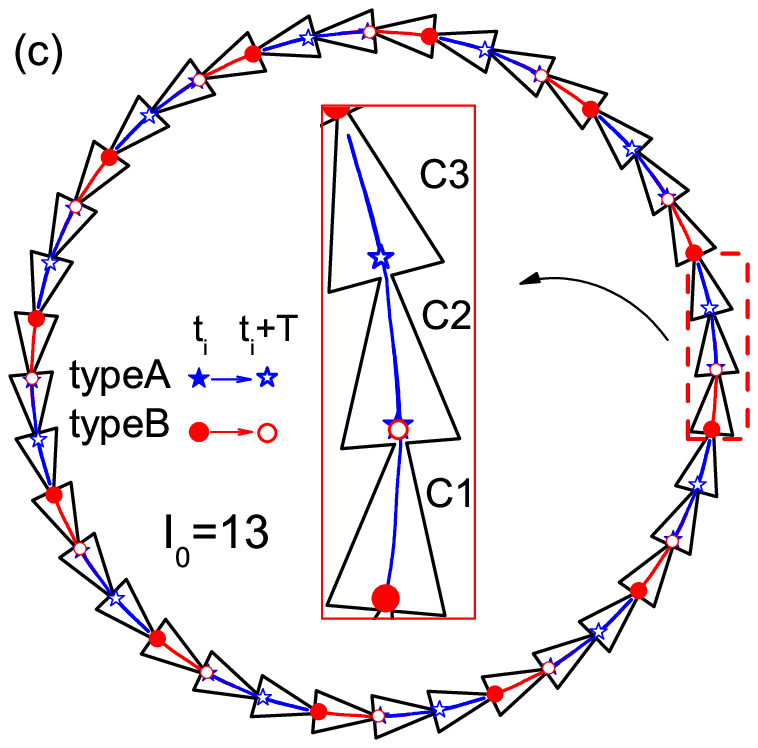}
\includegraphics[width=\textwidth]{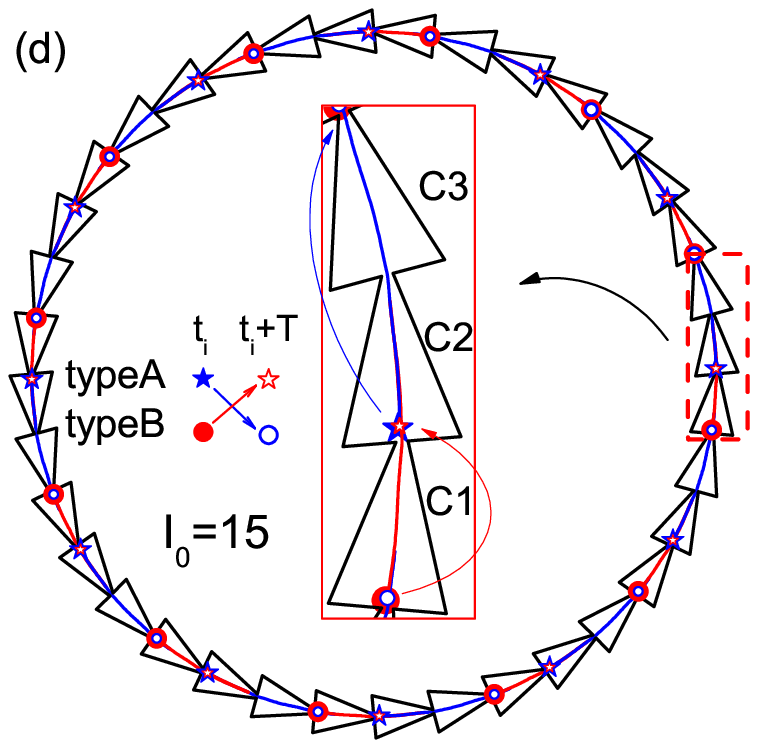}
\end{minipage}
\end{figure*}

As we defined above, the system contains $L$ vortices and $N$ triangles in a circular chain.
If there is a common integer (except one) in terms of which two numbers $L$ and $N$ can both be measured, then they are commensurate.
Otherwise, they are incommensurate.
Fig.~\ref{fig:iw_com}) shows the average angular velocity $\omega$ as a function of $I_{0}$ for different commensurate ratios.
The ascending part of the $\omega(I_0)$ curves (i.e., where $\omega$ versus $I_{0}$ increases) is stepwise.
Besides the large steps of angular velocity in the $\omega-I_0$ curve for $L=1$, which we refer to as ``integer steps", we also found smaller steps for some specific vorticities $L$
(shown in Fig.~\ref{fig:iw_com}).
If $L/N=k/m$ and $k\neq1$, where $k$ and $m$ are incommensurate
integers, the small steps can be found in the $\omega-I_0$ curves
[in Fig. \ref{fig:iw_com}(b)-(d)]. When $k=1$, we observe only
integer steps of the angular velocity $\omega(I_0)$ [in Fig.
\ref{fig:iw_com}(a)]. The difference in the angular velocity between
two adjacent steps is always $\omega_0$. If $k\neq1$, then we can
find a fractional step whose magnitude is $\omega_0/k$. For example,
for $k=2$ [shown in Fig. \ref{fig:iw_com}(b)], the smallest step is
$\omega_0/2$ in the system with 72 vortices (i.e., $L=72$), which is
a half of that for $k=1$ (e.g., when $L=36$). In the case of $k=1$,
a unit cell contains only one vortex. [The ``unit cell" (UC) is a
minimum repeatable set of TC(s) containing an integer number of
vortices. For example, for $L=1$ the UC is the entire channel with
one vortex, and for $L=36$ the UC is one TC with one vortex.] If one
vortex can overcome the potential barrier and move in the easy
direction, all the vortices can do so at the same time, i.e.,
collectively. This results in integer steps of the average angular
velocity $\omega_0$ in the $\omega-I_0$ curve. However, fractional
steps appear when there are more than one vortex in each UC. For
instance, a UC contains one TC with two vortices when $L=72$, and if
$L=24$, the UC is constructed by three TCs with two vortices. In
general case, the unit cell contains $m$ TCs with $k$ vortices
inside. If $k>1$, there are more than one vortex in the UC, and
those vortices are not equivalent, i.e., they are not located at the
equivalent position in the TC and/or they experience different
interactions with the boundary. Therefore, they move with different
angular velocities in each period. Let us take $k=2$, for example
(see Fig. \ref{fig:typeab}). As shown in Fig. \ref{fig:typeab}(a),
the vortices located in a TC that is followed by an empty TC in the
easy direction (type-A) can overcome the potential barriers in a
period of alternating current and move to its neighbor TC, while the
others (type-B) are still localized in its original TC. When
$I_0=10$ [shown in Fig. \ref{fig:typeab}(b)], type-A vortices move
from C2 to C3 but type-B vortices only oscillate in C1. In the view
of the whole circular channel, only a half of vortices (i.e., type-A
vortices) move to another TC while the other half of vortices (i.e.,
type-B vortices) do not move. After that, the type-B vortices are
located at similar position as type-A vortices in the previous
cycle, i.e., the previous type-B vortices now become type-A vortices
and they will move to next TC in the anticlockwise direction during
the next period. Therefore, when we calculate the average angular
velocity for all the vortices, the value is a half of the angular
velocity of type-A vortices, i.e., $\omega_0/2$. If the current
increases, every vortex can overcome the barriers and move to the
neighbor TC [e.g., as shown in Fig. \ref{fig:typeab}, type-A
vortices move from C2 to C3 and type-B vortices move from C1 to C2
when $I_0=13$]. Then average angular velocity becomes $\omega_0$
which is the same as that for $k=1$ when all vortices move to the
neighbor TC.

\subsection{Commensurability effect of frequency}

\begin{figure*}[tb]
\includegraphics[width=0.40\textwidth]{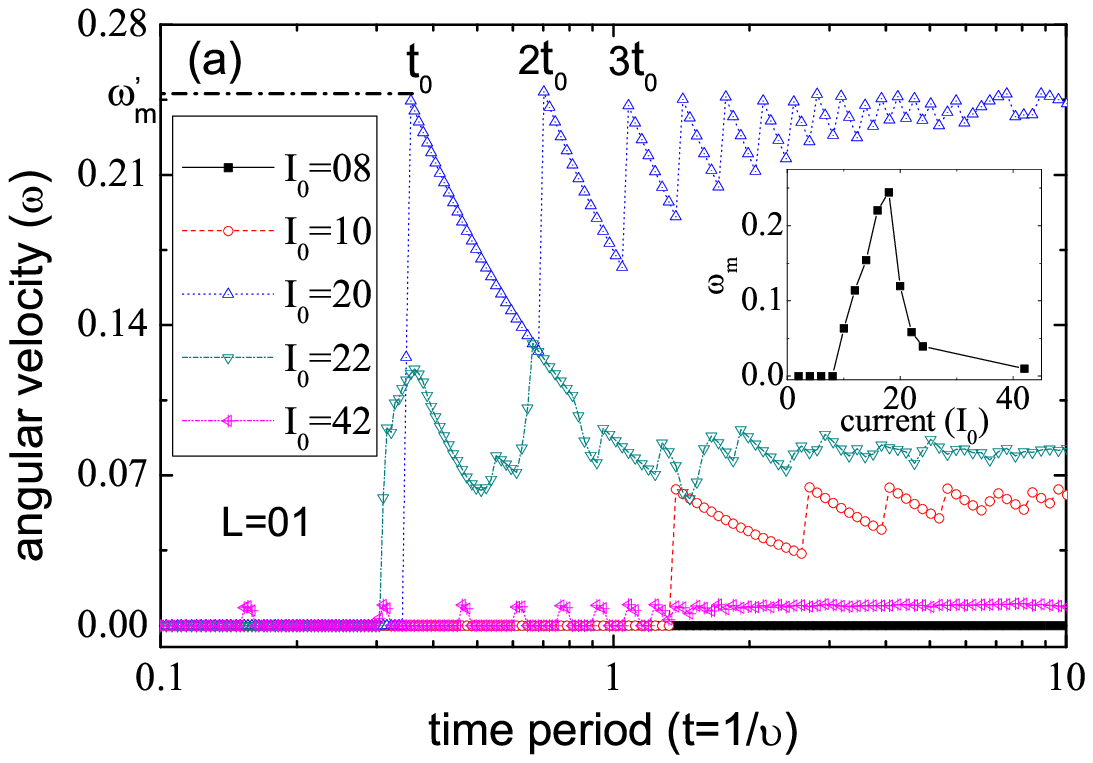}
\includegraphics[width=0.40\textwidth]{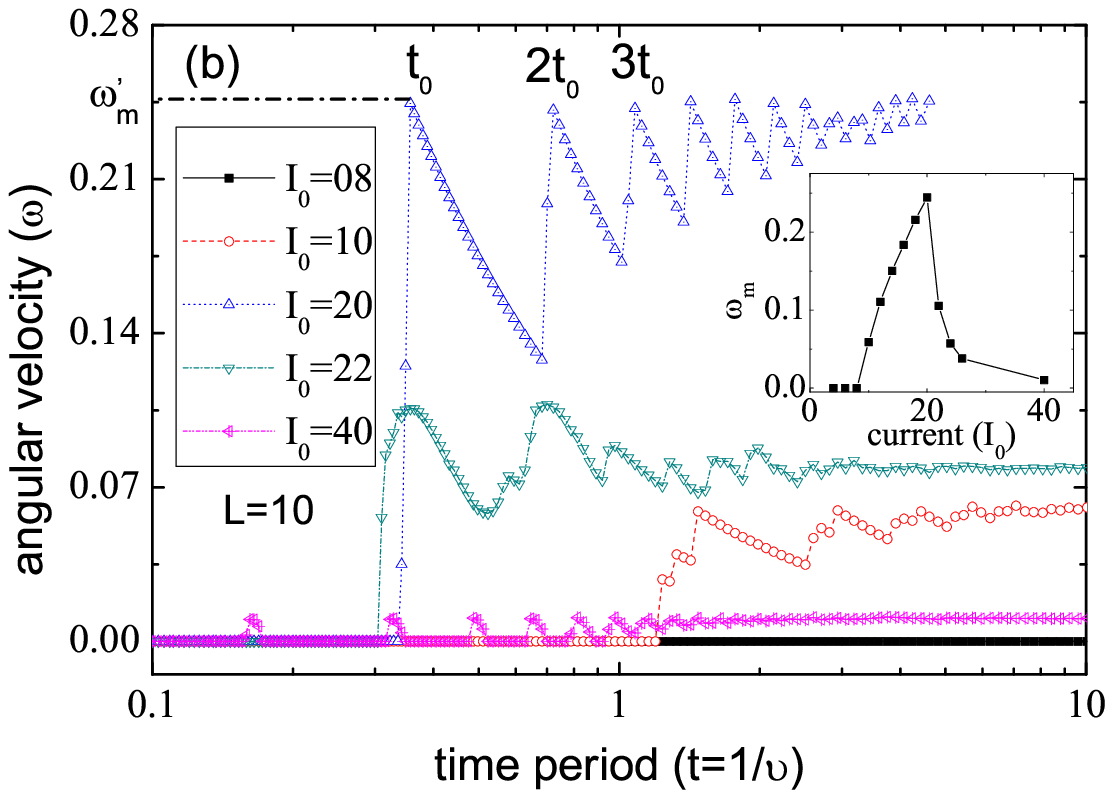}
\includegraphics[width=0.40\textwidth]{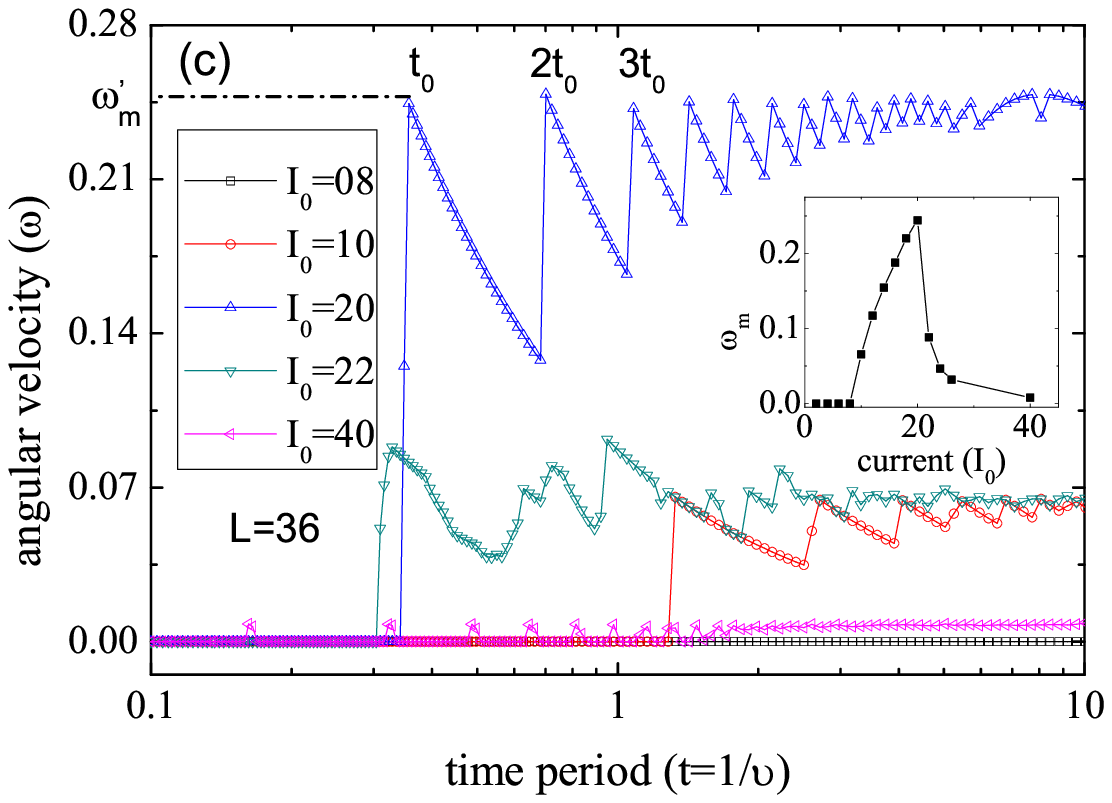}
\includegraphics[width=0.40\textwidth]{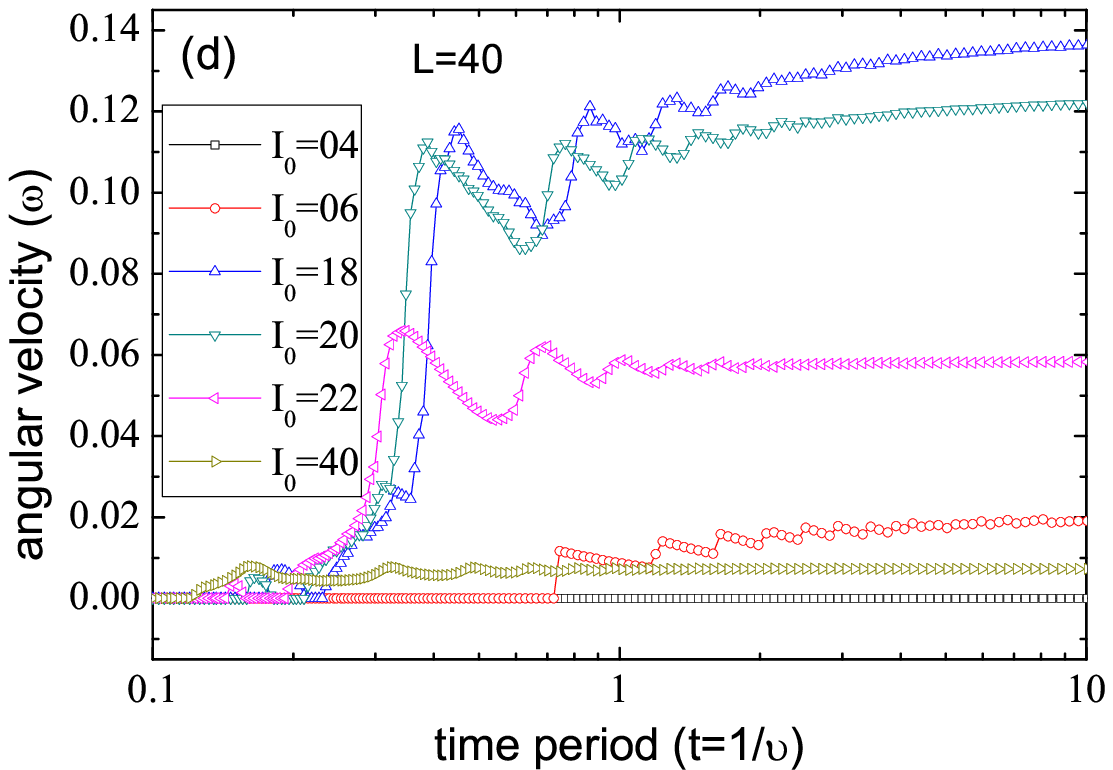}
\includegraphics[width=0.40\textwidth]{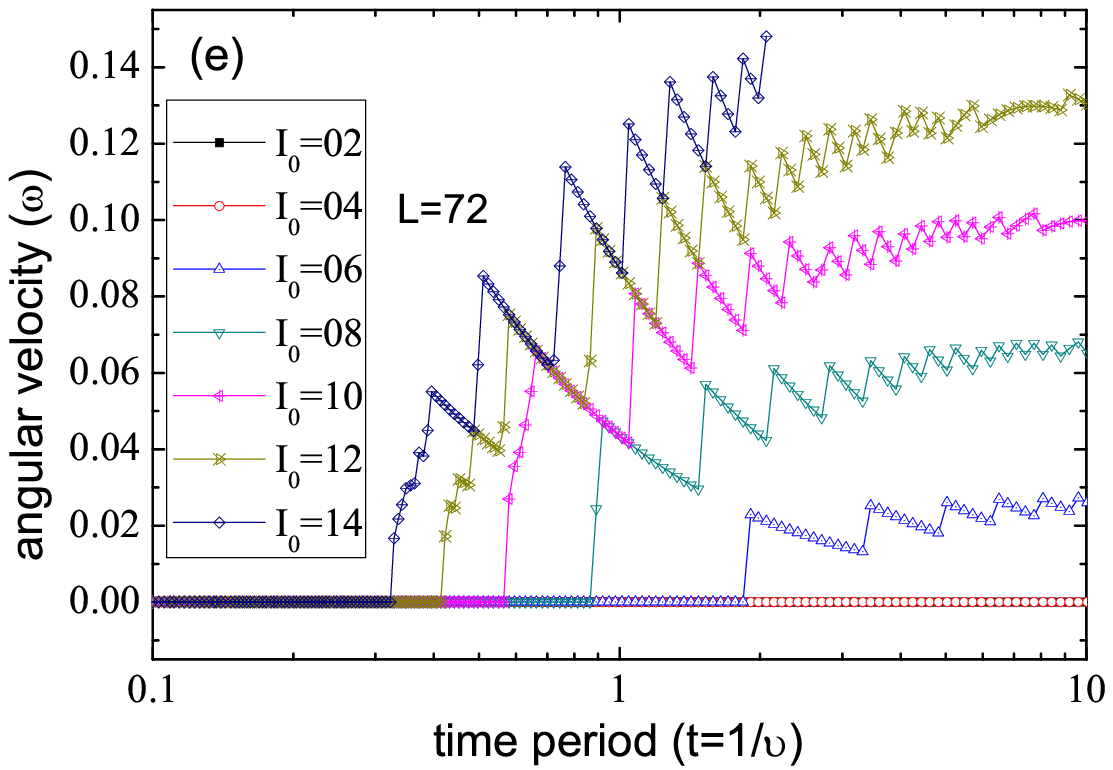}
\includegraphics[width=0.40\textwidth]{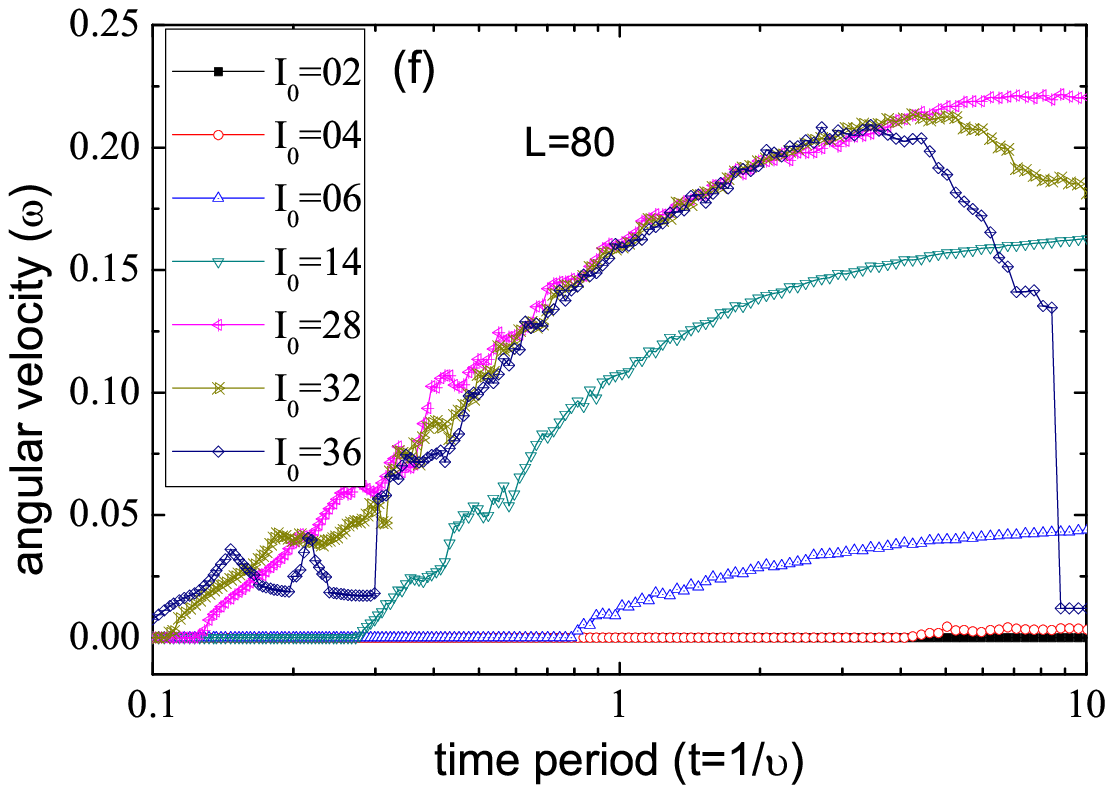}
\caption{
(Color online)
The $\omega-t$ curves
in a channel with $g = 0.1$
for different density of vortices:
$L/N\ll1$ (a), $L/N<1$ (b), $L/N=1$ (c) (commensurate), $2>L/N>1$
(d), $L/N=2$ (e) (commensurate), and $L/N>2$ (f). When the density
$L/N\leq1$, i.e., in the single-vortex regime, the angular velocity
$\omega(t)$ first reaches the maximum and then oscillates [see (a),
(b) and (c)]. With increasing current, the maximum velocity
increases until the driving force reaches the value larger than
$f_m^-$. For high density, the angular velocity $\omega(t)$ does not
reach the maximum because the vortex motion in the easy direction is
compensated by that in the hard direction. $\omega(t)$ increases
when the frequency decreases. In commensurate cases [(c) and (e)]
$\omega(t)$ oscillates with sharp jumps which are not observed in
incommensurate cases for high vortex density (f). }\label{fig:fw}
\end{figure*}

Further we analyze the angular velocity evolution while varying the ac drive frequency (shown in Fig. \ref{fig:fw}).
For varying frequency $\nu$, the $\omega-\nu$ curves are characterized by peaks and/or oscillations.
Let us clarify this behavior.
As discussed in Sec. III, if $f^-_m>f^d>f^+_m$, vortices can move in the easy direction and be frozen/blocked in the hard direction when the current alternates.
Let us introduce a time scale, $T_0$, to
characterize the motion of a vortex over the entire TC. One TC
occupies an angle $\theta_0=2\pi/N$. If we assume that a vortex
moves over the entire TC to the equivalent position in the next TC,
then $T_0$ should satisfy the following condition:
\begin{equation}\label{eq:t0}
    \int_{0}^{T_0}\omega dt= \int_{0}^{T_0}[\boldsymbol{f}^d(I_0)+
    \boldsymbol{f}^b+ \boldsymbol{f}^{vv}]\cdot\hat{\theta}dt/
    \eta r= \theta_0,
\end{equation}
where $\hat{\theta}$ is the unit vector in the azimuthal direction.
In the single vortex case ($L=1$), we set $\boldsymbol{f}^{vv}=0$,
and $\theta_0$ is a constant. Considering the circular trajectory of
the moving vortex, the integral of $\boldsymbol{f}^b$ is also a
constant. Therefore, $T_0$ only depends on the current $I_0$ and we
use the notation $T_0(I_0)$ instead of $T_0$ in order to show the
dependence on the driving current.

If $f^-_m>f^d(I_0)>f^+_m$ and the ac period $t=2T_0(I_0)=t_0(I_0)$,
then the rectified signal will be maximum since all the vortices
coherently move over the entire TC in the easy direction during the
first half period and do not move backwards in the next half period.
Then we can obtain the principal period $t_0(I_0)=2T_0(I_0)$ for
each value of the current and roughly estimate forces $f^-_m$ and
$f^+_m$ from the $\omega-\nu$ or $\omega-t$ curves [as shown in Fig.
\ref{fig:fw}]. The maximum angular velocity $\omega_m$ for different
current and vorticities $L$ is shown in the insets of Fig.
\ref{fig:fw}. For example, in the case of $L=1$ [see Fig.
\ref{fig:fw}(a)], the angular velocity is always zero when
$I_0\leq8$ and becomes non-zero for $I_0=10$, i.e., the vortex
starts to move in the easy direction. Therefore,
$f^d(I_0=8)<f^+_m<f^d(I_0=10)$. However, for $I_0\geq22$, the
maximum angular velocity $\omega_m$ decreases as compared to the
case for $I_0=20$, which means $f^d(I_0=20)<f^-_m<f^d(I_0=22)$. For
$f^d(I_0)>f^-_m$ when driving current increases, vortices move
backwards (i.e., the motion in the hard direction) therefore
resulting in the decreasing net angular velocity.
For example, when $I_0=22$ [shown in Fig. \ref{fig:fw}(a)], the maximum $\omega_m$ becomes smaller and the jumps in the $\omega-t$ curves become smoother, which means the effect of boundary becomes weaker under a strong driving force.
Therefore, the angular velocity should be zero when the driving force goes to infinity. It explains why the angular
velocity decreases and goes to zero when the driving current
increases above some critical value.

\begin{figure}
  \includegraphics[width=0.45\textwidth]{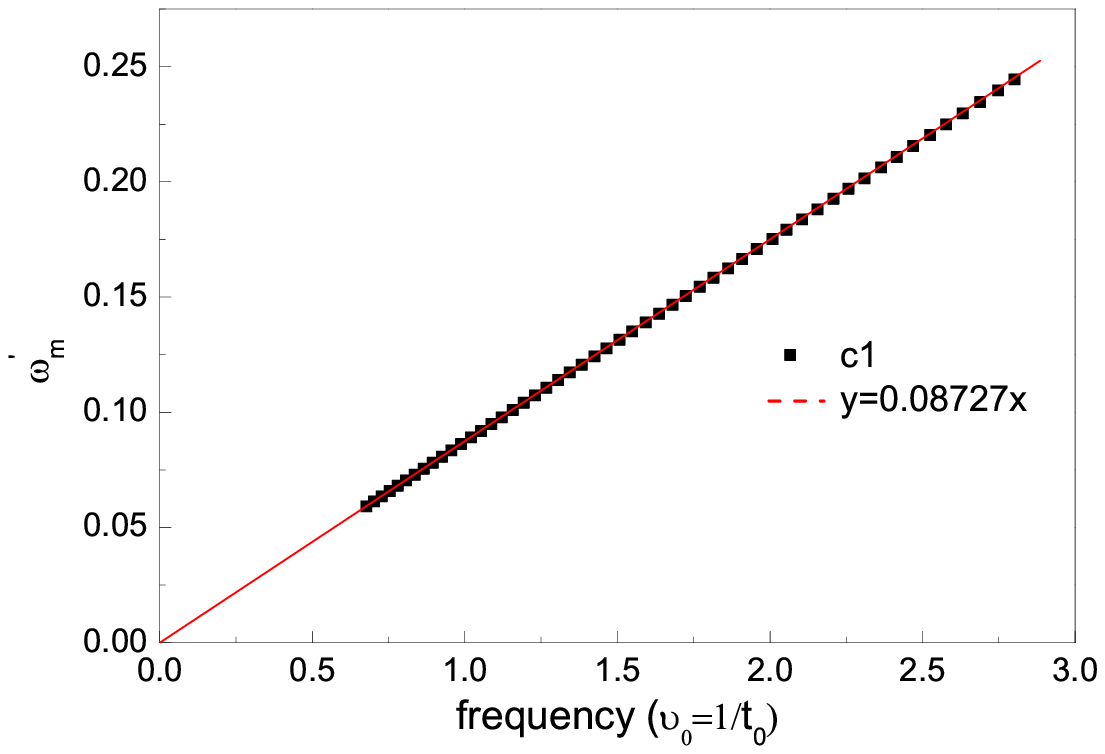}\\
  \caption{
(Color online)
The frequency dependence of the first local maximum
  $\omega_m^\prime$ of the $\omega-t$ curves for vortices in the single-vortex regime when
  the driving force is smaller than $f_m^-$. The magnitude of the
  first peak shows a linear dependence on the principal frequency $\nu_0$.}
  \label{fig:line}
\end{figure}

For the single-vortex case, the frequency dependence of the first
local maximum $\omega_m^\prime$ in the $\omega - \nu$ curves versus
the corresponding drive frequencies, $\nu_0 = 1/t_0$, is plotted in
Fig.~\ref{fig:line} which shows a linear behavior.
When the density is increased up to one vortex per cell ($L/N\leq 1$), the $\omega-t$ curve is similar to the one for $L=1$  for both
incommensurate [e.g., see Fig. \ref{fig:fw}(b)] and commensurate
cases [see Fig. \ref{fig:fw}(c)] because the interaction between
vortices is weak. For a higher density, the (in-)commensurability effect becomes pronounced.
The sharp jumps of the angular velocity are obtained only in the commensurate case [e.g., $L=72$ shown in
Fig. \ref{fig:fw}(e)] but not in the incommensurate case [e.g.,
$L=80$ shown in Fig. \ref{fig:fw}(f)]. Comparing the high density
cases with the low density cases, we can conclude that sharp jumps
in $\omega-t$ curve, which are both found in single-vortex and
multi-vortex regimes, are due to the effect of periodically repeated
boundaries of the ratchet potential and the increasing local maximum
of the angular velocity when $t$ increases in the multi-vortex
regime. This is different from the single-vortex regime, because of
the strong interaction between vortices induced by the high vortex
density. If the principal period (i.e., the time period for the
first jump) is denoted by $t_0$, the others will be harmonics of
$t_0$, $kt_0\ (k=2,3,...)$.

\section{Experimental detection of vortex ratchet effect in
a Corbino geometry}

\begin{figure}
  \includegraphics[width=0.45\textwidth]{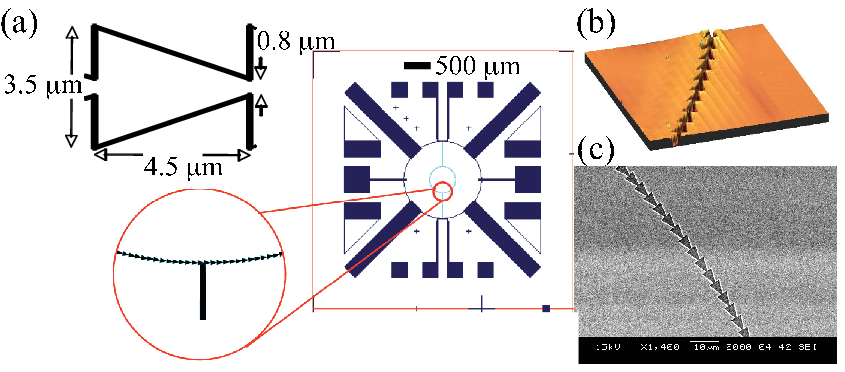}\\
\caption{
(Color online)
(a) Detail of single ratchet cell of channel and schematic of Corbino
ratchet chip layout; extra pads and leads beyond disk were not used
for the measurements presented here.
(b) Atomic force microscope image of portion of Corbino ratchet channel.
(c) Scanning electron micrograph of portion of Corbino ratchet channel.
}
\label{fig:exp01}
\end{figure}

We have performed preliminary measurements on a device containing
a single weak-pinning circular channel for guiding vortex motion
in a Corbino geometry.
This is based on a technique for using weak-pinning channels with
tailored edges to produce asymmetric vortex confining potentials.
Such an arrangement resulted in substantial asymmetric vortex
response for linear channels on a strip geometry
\cite{10PRB2007Kes}.
The Corbino sample consists of a Si substrate with a 200-nm thick
film of weak-pinning a-NbGe and a 50-nm thick film of strong-pinning
NbN on top.
The fabrication followed the scheme of previous weak-pinning channel
devices
\cite{10PRB2007Kes,Plourde2008},
with the 1.5-{\rm mm} diameter Corbino disk pattern etched through
the entire superconducting bilayer.
The 500-$\mu${\rm m} diameter circular chain of triangular cells
was etched through the NbN layer
(Fig.~\ref{fig:exp01}),
thus defining the
weak-pinning a-NbGe channel region for vortex flow.

Wirebonds were attached between the center and perimeter of the
Corbino disk for injecting a bias current with a radial flow.
Because of the rather small flux-flow voltages for vortex motion
in a single channel, it was necessary to use a custom picovoltmeter
based on a dc SQUID operated in a flux-locked loop
\cite{Plourde2008}.
Measurements of the noise power at different temperatures were used
to calibrate the value for the series resistance at the SQUID input,
thus allowing for a measurement of the system gain, as described in
Ref.~\cite{Plourde2008}.
During the measurements, the sample and SQUID were immersed in
a pumped liquid helium bath. Shielding of external magnetic fields
was achieved with a $\mu$-metal shield surrounding the dewar and
a superconducting Pb shield around the sample and SQUID on the
bottom of the cryogenic insert.
Vortices were introduced into the channel by temporarily raising
the sample above the helium bath, heating to
$\sim$6 {\rm K} -- between $T_c^{\rm NbGe}$ = 2.88 {\rm K}
and $T_c^{\rm NbN}$ $\approx$ 10 {\rm K} -- while applying a small
magnetic field with a Helmholtz coil on the insert, then cooling
back down below $T_c^{\rm NbGe}$.

\begin{figure}
  \includegraphics[width=0.45\textwidth]{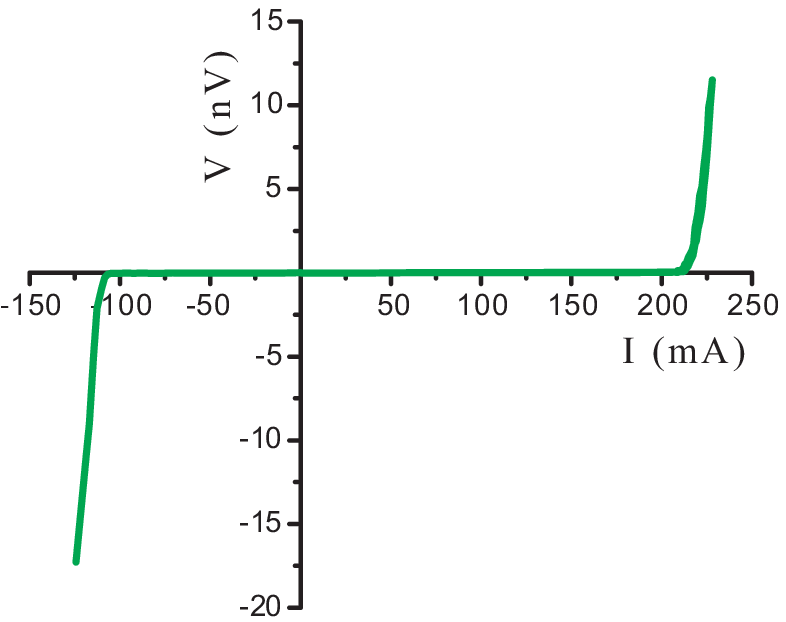}\\
\caption{
(Color online)
Current-voltage characteristic of Corbino ratchet channel cooled
to 1.60 {\rm K} in an external magnetic field of 0.26 {\rm Oe}
as described in text.
}
\label{fig:exp02}
\end{figure}

Upon reaching the desired measurement temperature, the bias current
was varied incrementally and the flux-flow voltage sensed by the
SQUID was recorded for each current value.
Such current-voltage characteristics (IVC) exhibited substantial
asymmetries between the critical current for vortices to begin
to move through the channel for the two directions.
This asymmetry persisted down to the lowest measurement temperature,
1.60 {\rm K}, well below the onset of superconductivity in the channel
at $T_c^{\rm NbGe}$
(Fig.~\ref{fig:exp02}).
The larger critical current corresponded to the sense of vortex motion
in the hard direction through the ratchet cells.
Due to experimental wiring limitations on these preliminary measurements
of a Corbino ratchet channel, it was not possible to sweep the bias current
with an oscillatory drive to study the average flux-flow voltage over a cycle.
Nonetheless, the large asymmetry between the two senses of critical current
demonstrates the potential for a weak-pinning ratchet channel in a Corbino
geometry to rectify vortex motion.


\section{Conclusions}

A vortex moving in an asymmetric circular channel in a Corbino setup
experiences the confinement of the boundary, the repulsive
interaction due to other vortices and the gradient (i.e., radially
decreasing) driving force when an external current is applied. The
combination of these factors determines the vortex motion. Different
dynamical behavior is observed in low and high density systems,
which are referred to as ``single-vortex" and ``multi-vortex"
regimes, respectively. For low density, i.e., in the single-vortex
regime, there is no more than one vortex per triangular cell so
that the vortex-vortex interaction is negligible. Therefore, the
ratchet potential due to the boundary dominates and all the vortices
follow circular 1D trajectories. Since all the vortices still move
in 1D when there is more than one vortex per cell (i.e., a higher density),
a vortex can escape even easier either in the easy or in the hard
direction due to the interaction between vortices in the same cell.
This results in decrease of the rectified net current.
However, when the number of
vortices increases further (i.e., in the``multi-vortex" regime of
rectification when the rectified net current increases with the
vortex density), the trajectories of vortex motion for low driving
currents are 2D while these trajectories squeeze and turn to 1D
circle with increasing driving current. Because of the circular
geometry of the channel, the density of vortices first becomes
inhomogeneous [i.e, see Fig. \ref{fig:trace_high}(b)] during the
transition from 2D motion to 1D motion in the multi-vortex regime
and then becomes homogeneous again when all the vortices move in a
circular trajectory.

Considering the asymmetry in the radial direction, vortices near the inner/outer corner of the triangular cells (TCs) (i.e., closer/further to the center of the disk) are driven by different Lorentz forces and for some specific value of driving
current the vortex in the outer corner moves to the inner corner
while the one in the inner corner moves to the next TC [e.g., see
Fig. \ref{fig:trace_high}(a)]. This kind of motion prevents vortices
from arriving simultaneously at the narrow part that would lead to
jamming which occurs in case of a linear channel. When the density
increases, the maximum net angular velocity $\omega_m$ remains the
same in the single-vortex regime and then decreases until reaching
the multi-vortex regime, and then $\omega_m$ increases.

The frequency of driving current also strongly influences the vortex
dynamical behavior. The ac frequency determines the possible distance
a vortex moves during an ac period. For high frequency, a nearly
zero net angular velocity is obtained for different values of
driving current $I_0$. Each vortex is unable to overcome the energy
barrier and is confined by a single potential well instead of the
periodic ratchet potential. When the frequency is low, the ratchet
effect is clearly observed in the $\omega-I_0$ curve but the
commensurability effect is not present. For an intermediate
frequency of driving current under which the distance of moving
vortex in a period is comparable to the size of the TC, both the
ratchet effect and the commensurability effect have been observed.

Besides the vortex density and the frequency of current, the
commensurability between the numbers of vortices and TCs also plays
an important role in the dynamical behavior of vortices, which leads
to jumps in the angular velocity $\omega$ with increasing driving
current $I_0$ (i.e., steps in the $\omega-I_0$ curve). Therefore,
under some specific conditions, the average angular velocity of
vortices is not a continuous function of the driving current.
The commensurability also influences the minimum difference of the
angular velocity for different steps in the $\omega-I_0$ curve, and
results in integer steps (i.e., the large steps which are found,
e.g., for vorticity $L=1$) and fractional steps (i.e., the smaller
steps whose magnitude are fractions of the magnitude of the integer
steps, e.g., for vorticity $L=24$) in a certain range of the
current. We also obtained several peaks (sharp jumps) in the
$\omega-t$ curve, which correspond to the principal period $t_0$
(during which the vortex can move over one TC in the easy
direction), and harmonics periods $k\nu_0\ (k=2,3,...)$ (during
which vortices move over $k$ TCs in the easy direction). The net
flow of vortices is enhanced when the ac period is one of the
harmonics periods, i.e., the average angular velocity reaches a local
maximum.

\section*{Acknowledgement}
We thank Peter Kes and Marcel Hesselberth for providing the superconducting films from which the Corbino ratchet sample was fabricated.
This work was supported by the ``Odysseus'' Program of the Flemish
Government and the Flemish Science Foundation (FWO-Vl), the
Interuniversity Attraction Poles (IAP) Programme --- Belgian State
--- Belgian Science Policy, and the FWO-Vl (Belgium).
T.W.H., K.Y., and B.L.T.P acknowledge support from the
National Science Foundation under Grant DMR-0547147,
as well as the use of the Cornell NanoScale Facility,
a member of the National Nanotechnology Infrastructure Network,
which is supported by the National Science Foundation (Grant ECS-0335765).


\begin{thebibliography}{00}


\bibitem{SMadd1}E. R. Dufresne, D. Altman, and D. G. Grier,
Europhys. Lett. \textbf{53}, 264 (2001); E. R. Dufresne, T. M.
Squires, M. P. Brenner, and D. G. Grier, Phys. Rev. Lett.
\textbf{85}, 3317 (2000).

\bibitem{SMadd2} R. D. Astumian, Phys. Rev. Lett. \textbf{91},
 118102 (2003).

\bibitem{2PRB2005Peeters}
D.Y. Vodolazov and F.M. Peeters, Phys. Rev. B \textbf{72}, 172508
(2005).

\bibitem{7PRB2005Nori}
S. Savel'ev, V. Misko, F. Marchesoni, and F. Nori, Phys. Rev. B
\textbf{71}, 214303 (2005).

\bibitem{mod}
P. H\"{a}nggi and F. Marchesoni, Rev. of Mod. Phys. \textbf{81}, 387
(2009).

\bibitem{10PRB2007Kes}
K. Yu, T.W. Heitmann, C. Song, M. P. DeFeo, B.L.T. Plourde, M.B.S.
Hesselberth, and P.H. Kes, Phys. Rev. B \textbf{76}, 220507 (2007).

\bibitem{13PRL1999Nori}
J.F. Wambaugh, C. Reichhardt, C.J. Olson, F. Marchesoni, and F.
Nori, Phys. Rev. Lett. \textbf{83}, 5106 (1999).


\bibitem{IEEE2009Plourde}
B.L.T. Plourde, IEEE Trans. Appl. Supercon., \textbf{19}, 3698
(2009).

\bibitem{3Nature2006}
 C.C. de Souza Silva, J. Van de Vondel, M. Morelle, and V.V.
 Moshchalkov, Nature \textbf{440}, 651-654 (2006).

\bibitem{4Scinece2003}
J.E. Villegas, S. Savel'ev, F. Nori, E.M. Gonzalez, J.V. Anguita, R.
Garc\'{i}a, and J.L. Vicent, Science \textbf{302}, 1188 (2003).

\bibitem{8PRB2005Vincent}
J.E. Villegas, E.M. Gonzalez, M.P. Gonzalez, J.V. Anguita, and J.L.
Vicent, Phys. Rev. B \textbf{71}, 024519 (2005).

\bibitem{9PRL2005Moshchalkov}
J. Van de Vondel, C.C. de Souza Silva, B.Y. Zhu, M. Morelle, and
V.V. Moshchalkov, Phys. Rev. Lett. \textbf{94}, 057003 (2005).

\bibitem{12PRB2010Mashchalkov}
B.B. Jin, B.Y. Zhu, R. W\"{o}rdenweber, C.C. de Souza Silva, P.H.
Wu, and V.V. Moshchalkov, Phys. Rev. B \textbf{81}, 174505 (2010).

\bibitem{highTC2004} R. W\"{o}rdenweber, P. Dymashevski, and V. R. Misko,
Phys. Rev. B \textbf{69}, 184504 (2004).

\bibitem{PRL2005Tonomura} Y. Togawa, K.
Harada, T. Akashi, H. Kasai, T. Matsuda, F. Nori, A. Maeda, and A.
Tonomura, Phys. Rev. Lett. \textbf{95}, 087002 (2005).

\bibitem{6PRL2001Nori}
C.J. Olson, C. Reichhardt, B. Jank\'{o}, and F. Nori, Phys. Rev.
Lett. \textbf{87}, 177002 (2001).

\bibitem{16PRL2007Reichhardt}
W. Gillijns, A.V. Silhanek, V.V. Moshchalkov, C.J. Olson Reichhardt,
and C. Reichhardt, Phys. Rev. Lett. \textbf{99}, 247002 (2007).

\bibitem{15ref1Nori}P. H\"{a}nggi, F. Marchesoni, and F. Nori, Ann.
Phys. (Leipz.) \textbf{14}, 51-70 (2005).

\bibitem{15ref2Marchesoni}P. H\"{a}nggi and F. Marchesoni, Chaos
\textbf{15}, 026101 (2005).

\bibitem{15NMat2006Nori}
D. Cole, S. Bending, S. Savel'ev, A. Grigorenko, T. Tamegai, and F.
Nori, Nature Materials \textbf{5}, 305-311 (2006).

\bibitem{Nori_Reichhardt} C. Reichhardt, C.J. Olson, and F. Nori,
Phys. Rev. Lett. \textbf{78}, 2648 (1997); Phys. Rev. B \textbf{57},
7937 (1998); Phys. Rev. B \textbf{58}, 6534 (1998).

\bibitem{PRL2006PRB2007misko}V.R. Misko, S. Savel'ev, A.L.  Rakhmanov, and F. Nori,
Phys. Rev. Lett. \textbf{96}, 127004 (2006); Phys. Rev. B
\textbf{75}, 024509 (2007).

\bibitem{1PRB2010Reichhardt}
C.J. Olson Reichhardt and C. Reichhardt, Phys. Rev. B \textbf{81},
224516 (2010).

\bibitem{14PhysC2010Reichhardt}
C. Reichhardt and C.J. Olson Reichhardt, Physica C (2010),
doi:10.1016/j.physc.2010.02.067.

\bibitem{17PRB2007Reichhardt}
Q. Lu,  C.J. Olson Reichhardt, and C. Reichhardt, Phys. Rev. B
\textbf{75}, 054502 (2007).

\bibitem{Nature1999}
C.S. Lee, B. Jank\'{o}, I. Der\'{e}nyi, and A.L. Barab\'{a}si,
Nature \textbf{400}, 337-340 (1999).


\bibitem{11PRB2010Plourde}
K. Yu, M.B.S. Hesselberth, P.H. Kes, and B.L.T. Plourde, Phys. Rev.
B \textbf{81}, 184503 (2010).


\bibitem{TransRatchetAPL2007}
E.M. Gonzalez, N.O. Nunez, J.V. Anguita, and J.L. Vicent, Appl.
Phys. Lett. \textbf{91}, 062505 (2007).

\bibitem{TransRatchetAPL2008}
A.V. Silhanek, J. Van de Vondel, V.V. Moshchalkov, A. Leo, V.
Metlushko, B. Ilic, V. R. Misko, and F. M. Peeters, Appl. Phys.
Lett. \textbf{92}, 176101 (2008).

\bibitem{Plourde2008} T.W. Heitmann, K. Yu, C. Song, M.P. DeFeo,
 B.L.T. Plourde, M.B.S. Hesselberth, and P.H. Kes,
 Rev. Sci. Instrum. \textbf{79}, 103906 (2008).

\bibitem{DynamicHTSC} D. L\'{o}pez, W. K. Kwok, H. Safar,
R.J. Olsson, A. M. Petrean, L. Paulius, and G. W. Crabtree, Phys.
Rev. Lett. \textbf{82}, 1277 (1999).

\bibitem{Crabtree1999} G. W. Crabtree, D. L\'{o}pez, W. K. Kwok, H. Safar, and
L. M. Paulius, J. Low Temp. Phys. \textbf{117}, 1313 (1999).

\bibitem{lgCorbinoMiguel} M. C. Miguel and S. Zapperi,
Nature Materials \textbf{2}, 477 (2003).

\bibitem{Marchetti2002}P. Benetatos and M. C. Marchetti, Phys. Rev.
B \textbf{65}, 134517 (2002).

\bibitem{smCorbinoMisko}V. R. Misko and F. M. Peeters,
Phys. Rev. B \textbf{74}, 174507 (2006).

\bibitem{PRLLinMisko}
N. S. Lin, V. R. Misko, and F. M. Peeters,
Phys. Rev. Lett. \textbf{102}, 197003 (2009); Phys. Rev. B
\textbf{81}, 134504 (2010); V. R. Misko, N. S. Lin, and F. M.
Peeters, Physica C \textbf{47}, 939 (2010).

\bibitem{fvvBae}B.J. Baelus, L.R.E. Cabral, and F.M. Peeters,
Phys. Rev. B \textbf{69}, 064506 (2004).

\bibitem{Grigorieva2007} 
I.V. Grigorieva, W. Escoffier, V.R. Misko, 
B.J. Baelus, F.M. Peeters, L.Y. Vinnikov, and S.V. Dubonos, Phys. Rev. Lett. \textbf{99}, 147003 (2007).

\bibitem{Bardeen65}
J. Bardeen and M.J. Stephen, Phys. Rev. {\bf 140}, A1197 (1965). 

\bibitem{babic} 
D. Babic, J. Bentner, C. Surgers, C. Strunk, Phys. Rev. B {\bf 69}, 092510 (2004). 

\bibitem{LO} 
A.I.~Larkin and Y.N.~Ovchinnikov, Zh. Eksp. Teor. Fiz. {\bf 68}, 1915 (1975).

\bibitem{klein} 
W. Klein, R.P.~Huebener, S. Gauss, J. Parisi, J. Low Temp. Phys. {\bf 61}, 413 (1985). 



\end{thebibliography}
\end{document}